\documentclass{aa}
\usepackage{natbib}
\bibpunct{(}{)}{;}{a}{}{,}
\usepackage{lscape,graphicx,graphics}

\def\arcsec{\hbox{$^{\prime\prime}$}}
\def\micron{\hbox{$\mu$m}}
\def\mn{\hbox{$\mu$m}}

\def\iso{{\em ISO}}
\def\sws{{\em ISO-SWS}}

\def\ir{infrared}
\def\mir{mid-infrared}

\def\fsl{fine structure line}
\def\hrl{H-recombination line}
\def\n{NGC }
\def\sf{star formation}
\def\wr{Wolf-Rayet}

\def\hii{H\,II}
\def\neii{[Ne\,II]}
\def\neiii{[Ne\,III]}

\def\siii{[S\,III]}
\def\siv{[S\,IV]}

\begin{document}

   \title{A mid-infrared spectroscopic survey of starburst galaxies: 
   excitation and abundances}

   \author{Aprajita Verma
          \inst{1}
          \and
	Dieter Lutz
          \inst{1}
	  \and
	Eckhard Sturm 
	\inst{1}
        \and  
	Amiel Sternberg	
	\inst{2}
        \and  
	Reinhard Genzel
	\inst{1}
        \and  
	William Vacca
	\inst{1}
          }

   \authorrunning{Verma, Lutz et al.}

   \titlerunning{A mid-infrared spectroscopic survey of starburst galaxies}

   \offprints{A. Verma, \email{verma@mpe.mpg.de}}

   \institute{Max-Planck-Institut f\"ur extraterrestrische Physik,
Postfach 1312, D-85741 Garching-bei-M\"unchen, Germany.
\email{verma@mpe.mpg.de, lutz@mpe.mpg.de, sturm@mpe.mpg.de, vacca@mpe.mpg.de}
         \and
             School of Physics and Astronomy, Wise Observatory, Raymond
and Beverly Sackler Faculty of Exact Sciences, Tel Aviv University.
Ramat Aviv, Tel Aviv 69978, Israel. \email{amiel@wise1.tau.ac.il}}

   \date{Received; accepted }

   \abstract{
        We present spectroscopy of mid-infrared emission
        lines in twelve starburst regions, 
	located in eleven starburst galaxies,
        for which a significant number of lines between 2.38 and 45$\mu m$
        were observed with the ISO Short Wavelength Spectrometer, with
        the intention of providing a reference resource 
	for mid-infrared spectra
        of starburst galaxies. The observation apertures were
        centred on actively star forming regions, including those
        which are inaccessible at optical
        wavelengths due to high levels of obscuration.
        We use this data set, which includes fine structure and
        hydrogen recombination lines, to investigate excitation
        and to derive gas phase abundances of neon, argon,
        and sulphur of the starburst galaxies.
        The derived Ne abundances span 
	approximately an order of magnitude, 
        up to values of $\sim3$ times solar.  
        The excitation ratios measured from the Ne and Ar lines 
	correlate well with
        each other (positively) and with abundances (negatively). 
	Both in excitation and abundance, a separation of objects
        with visible \wr\ features (high excitation, low abundance) is
        noted from those without (low excitation, high abundance). 
        For a given abundance, the starbursts are of relatively
        lower excitation than a comparative sample of H\,II
        regions, possibly due to ageing stellar populations.  
	By considering the abundance ratios of S with Ne and Ar we
        find that, in our higher metallicity systems, S is relatively
        underabundant by
        a factor of $\sim 3$.  We discuss the origin
        of this deficit and favour depletion of S
        onto dust grains as a likely explanation.  This
        weakness of the \mir\ fine structure lines of sulphur
        has ramifications for future infrared missions such as
        {\em SIRTF} and {\em Herschel} since it
        indicates that the S lines are less favourable tracers of star
        formation than is suggested by nebular models which do
        not consider this effect.
  
	In a related
        paper \citep{stu02}, we combine our results with spectra
        of Seyfert galaxies in order to derive diagnostic diagrams
        which can effectively discriminate between the two types of activity
        in obscured regions on the basis
        of excitation derived from detected mid-infrared lines.

   }

   \maketitle


\section{Introduction}

By definition, starburst galaxies are hosts to sites of recent star
formation with associated rates that cannot be maintained over a Hubble
time, which can be much higher than that determined for the Milky
Way. Starburst galaxies are believed to contribute significantly to the 
population of massive stars in the Universe
\citep{gal95}. Within 10Mpc, approximately 25\% of high
mass star formation is attributed to only four starburst galaxies
\citep{hec98}.  The reported increase in the star 
formation density of the Universe for $0.1 \la z \la 2$
\citep[e.g.][]{mad96,mad98,bar00}
and the high frequency of starbursts with
disturbed morphologies or in interacting/merging systems both imply that
starbursts are likely to play an important role in galaxy formation 
and evolution scenarios.
By virtue of their proximity, local starbursts can be used 
as astrophysical laboratories to investigate the processes that 
may be ongoing at higher redshifts. 
However, despite intensive observational and theoretical
modelling campaigns, the properties of local starbursts are still
not fully understood.  In particular, their stellar populations and 
the state of the interstellar medium remain difficult to constrain.

Observable expressions of a starburst depend upon a number of factors,
including the stellar initial mass function (IMF), star formation
history and the ageing/evolution of the stellar population. Both 
the evolution of the stellar population and the spectra of  
individual stars are a function of metallicity, which provides
the motivation for studying the effects of metallicity that we
consider in this paper. A galaxy's interstellar medium 
affects 
nebular diagnostics and, through obscuration, also direct stellar ones. 
The widely used local \citet{sca86} IMF may not be
reflective of the stellar mass content for starburst galaxies. IMFs modified
at either the low or high mass end have been invoked to explain 
observationally derived properties of starbursts such as dynamical
masses, spectral line strengths, continuum flux and radial velocities
\citep[e.g.][]{rie80,pux89,doy94,doh95,ach95,beck97,ste98,coz01}.
The limit on the upper end of the IMF
remains a controversial issue; mass limits as low as
$30M_{\sun}$ have been proposed. However, the presence of high mass
stars (up to $100M_{\sun}$) is well known in sites of extreme star formation located both in the
Galaxy and the local universe \citep[e.g.][]{kra95,ser98,
eis98,mas98,tre01}. In addition, the detection of
\wr\ features in several starbursts implies that their progenitor 
stars with masses $\ga25-60M_{\sun}$ were once present
\citep[][ and references therein]{sch01}.  

Analysis of mid-infrared nebular emission lines shows that the 
excitation of dusty starbursts is often lower than that of Galactic \hii\
regions \citep{tho00}. From theoretical modelling of the
[NeIII]15.5\mn/[NeII]12.8\mn\ excitation ratio, \citet{tho00} suggested that
the stellar SEDs of 27 starburst galaxies were consistent with the
formation of massive stars ($50-100M_{\sun}$) and argued against
IMFs with upper
mass cut-offs lower than $30M_{\sun}$.  The low excitation
of galaxies that were originally forming high mass stars is then mainly
attributed to ageing
of the stellar population, as opposed to a low upper mass
limit to the IMF.
In addition,
the dependence of both stellar evolutionary tracks and nebular properties
(including excitation) on metallicity \citep[e.g.][]{bre99,tho00,giv02} 
suggests that elemental abundances must be considered in the analysis of
the issues detailed above. 

To date, abundance and excitation studies have been  
based primarily on optical and near-infrared data
\citep[e.g. ][]{olo95,kob96,coz99,consi00}, mainly due to
observational limitations from the ground. 
Optical abundance studies however do not reach the obscured 
regions dominating the activity of many infrared-selected starbursts.  
Even for starbursts which have been well-studied in the optical, a combination of radio and 
infrared measurements has conclusively demonstrated that some of the most 
active star forming sites are optically obscured, with most of the
bolometric luminosity emerging in the IR \citep[e.g.][]{gor01,vac02}. Thus, 
investigations of the densest 
regions of \sf\ are often restricted to the \ir\ due to extinction 
by dust at shorter wavelengths.

In general, optical- versus infrared-selected samples of starbursts
are sensitive to
different starburst properties. Star formation traced by optical observations 
is generally located within a disk having low to moderate extinction
with relatively low \sf\ rates ($<20M_{\sun}\
yr^{-1}$), while in
the \ir, the densest regions of star formation can be probed, occurring as
compact events with star formation rates which may reach $\la few \times 10^2
M_{\sun}\ yr^{-1}$ \citep{rie01}.
Infrared spectroscopy, probing the dominant obscured regions of objects
at the dusty end of this sequence, is needed for a complete understanding
of the local starburst population with implications for the extensive
populations of starbursts at higher redshift detected both in the UV/optical 
and infrared. 

Here, we present Infrared Space Observatory (\iso) Short
Wavelength Spectrometer (\sws) spectra (between 2.38 and 45$\mu m$) for 12
starburst regions observed in significantly more detail than most of the
galaxies studied by \citet{tho00}, a sample with which it partially
overlaps. This paper complements the numerous 
existing abundance analyses in the optical/near-infrared with
the first comparable \mir\ systematic study of a comprehensive sample
of starburst galaxies. 
Our analysis is focused upon the gas phase abundances 
of neon, argon and sulphur in this group of local starburst galaxies.
These elements emit the strongest emission lines 
in the mid-infrared spectra
of starbursts.
Deriving abundances from a combination of infrared fine structure
and hydrogen recombination lines
has a 
number of technical advantages that minimise uncertainties. 
Extinction in this wavelength range is low, and the derived abundances
(relative to H) are moderately insensitive
to variations in electron temperature and density 
[abundances
are $\propto T_e^{-0.7}$ \citep{givsm02} and the dependence on $n_e$
cancels for abundances relative to H, see Sect. 4.4.1. See also
discussion in \citet{nfs01}.]. 
For all three elements,
the major ionisation stages found in \hii\ regions are covered by the \sws\ 
spectra, which minimises the required ionisation correction factors.
We relate the results to nebular excitation and thereby probe the hot stellar 
population. 

The data presented include fainter transitions not included in the
abundance analysis, and will be useful as a reference for work with
upcoming mid-infrared spectrographs on 8m class telescopes as well as 
observations of fainter and/or higher redshift sources with
forthcoming infrared telescopes such as {\em SIRTF}, {\em SOFIA},  and 
{\em Herschel}.

The layout of this paper is as follows: In Sect. 2 we describe the
observations, the sample selection and data reduction.  In
Sect. 3 we present the line lists and in Sect. 4 the excitation
and calculated abundance analysis.  We discuss the results in Sect.
5 and finally, our conclusions
are presented in Sect. 6.


\section{Observations}

\subsection{Sample Selection}

Our sample consists of twelve regions in eleven galaxies which 
exhibit starburst
characteristics in the \ir.  Of the starburst galaxy observations by 
\sws\ present in the \iso\ Data Archive, we selected those
for which, in addition to fine structure lines, at least one hydrogen
recombination line was detected. 
This is a pre-requisite since a \hrl\
is used as the reference line for
the abundance estimation. As \mir\ \hrl s are faint in
extragalactic sources, this requirement therefore also restricts
our sample to local systems ($v_{med}=787 km s^{-1}$,
$v_{max}=3124 km s^{-1}$). We note that our sample was
neither homogeneously selected nor is it complete. Therefore, we investigate
each galaxy on an individual basis and analyse trends found for
this ensemble of starbursts.
Details of the starburst sample are given in Table \ref{tabsam}.

\begin{table*}
	\caption{Coordinates,
	redshifts and properties of the starbursts comprising our sample.
	\label{tabsam}}
	\vspace{0.5cm}
\begin{minipage}{18cm}

\begin{tabular}{lllrlllll}
\hline\hline
Source& 
Coordinates\footnote{of SWS pointing centre} &
Redshift& 
$D_L$\footnote{Luminosity distance calculated assuming $H_0 = 75 km s^{-1} Mpc^{-1}, \Omega_{0}=1$}&
$L_{IR}$\footnote{Infrared luminosity calculated using the prescription of \citet{hel85}}& 
Spec. Type& 
&
Notes\footnote{WR: \wr\ features. Further details from \citet{scha99} \&
references therein; BCD: Blue Compact Dwarf galaxy; }\\

	& 
(J2000)			&		
& 
(Mpc)&
($L_{\sun}$)&
&
&
 \\
	
\hline

{\bf NGC 253}	& 
00 47 33.20 $-$25 17 17.2	& 
0.000817 	&
3.5&
$1.69\times10^{10}$&
HII&
&
Sb/Sc, edge-on \\
		
{\bf IC 342}	& 
03 46 48.10 +68 05 46.8 	& 
0.000113 	&		
3.6&
$3.74\times10^{7}$&
HII&
&
Scd, face-on \\

{\bf II Zw 40}& 
05 55 42.60 +03 23 31.5	& 
0.002632 	&
7&
$2.04\times10^{9}$&
HII&
&
BCD, WR\\

{\bf M82}	& 
09 55 50.70 +69 40 44.4	& 
0.000677 	&
3.3&
$1.52\times10^{10}$&
HII&
&
Irr\\

{\bf NGC 3256}& 
10 27 51.20 $-$43 54 13.5	& 
0.009487 	&
36.6&
$2.59\times10^{11}$&
HII&
&
colliding pair\\

{\bf NGC 3690A} & 
11 28 33.60 +58 33 46.6 	& 
0.010411 	&
40.&
$3.40\times10^{11}$&
HII&
&
pec. merger, WR\\

{\bf NGC 3690B/C} & 
11 28 30.90 +58 33 45.6	& 
0.010447 	&
40.&
$3.42\times10^{11}$&
HII&
&
pec. merger, WR\\

{\bf NGC 4038/9}\footnote{Overlap region containing giant molecular
clouds 3, 4 and 5 from \citet{wil00}}& 
12 01 52.90 $-$18 53 04.0		& 
0.005477 	&
21.0&
$4.45\times10^{10}$&
HII&
&
interacting\\

{\bf NGC 4945}& 
13 05 27.40 $-$49 28 05.5	& 
0.001868 	&
4&
$4.52\times10^{10}$&
HII, Sy2\footnote{AGN contribution to mid-IR emission lines negligible 
                 \citep{spo00}}&
&
Sc, edge-on\\

{\bf NGC 5236}& 
13 37 00.50 $-$29 51 55.3	& 
0.001721 	&
5.0&
$1.29\times10^{10}$&
HII&
&
SAB(s), face-on, WR\footnote{but refer to Sects. \ref{survey} \& \ref{wrsect}}\\

{\bf NGC 5253}& 
13 39 55.80 $-$31 38 31.0& 
0.001348 	&
4.1&
$1.64\times10^{9}$&
HII&
&
Im pec, BCD, WR\\

{\bf NGC 7552}& 
23 16 10.80 $-$42 35 04.2	& 
0.005287 	&
21.2&
$6.73\times10^{10}$&
HII, LINER&
&
SBbc\\
\hline

\end{tabular}
\end{minipage}
\end{table*}

\subsection{Observations and Data Reduction}

The \sws\ instrument on-board \iso\ performed spectroscopic
measurements between 2.38 to 45.2\micron\ with grating
spectrometers
producing spectra of medium to high
resolution [{\em R=1000-2000}, see
\citet{degr96,lee02} for more details].  Within its operational 
wavelength range lie a number of previously seldom-observed emission 
lines including \fsl\ and \hrl s from which elemental
abundances may be determined.

The \iso\ Data Archive was used to collate 
{\em all} of the SWS grating spectra available for each galaxy.  
Most data sets
were taken using the SWS02 observation template, which provides 
medium resolution measurements of targeted lines.  In some cases, full
grating scans producing lower resolution spectra were taken (the SWS01
observations template) and a handful of medium resolution 
SWS06\footnote{Refer to
\citet{lee02} for details regarding the \sws\ and its observation modes.}
measurements were also made.  Where multiple observations were available,
care was taken not to combine data taken with different aperture
centres to ensure the same part of the galaxy was being measured for
each observation.  

We used the SWS interactive analysis (IA3\footnote{IA3 is a joint 
development of the SWS consortium. Contributing institutes 
are SRON, MPE, KUL and the ESA
Astrophysics Division. A standalone version of IA3 (OSIA) 
is available from http://sws.ster.kuleuven.ac.be/osia/.}) 
to reduce the data sets from the edited raw data
products.  The motivation for using IA3 to re-reduce the 
data (rather than the pipeline
reduction) was to enhance the quality of the reduced product by using
interactive tools for dark current subtraction, correction between up
and down scans and outlier masking.  The resultant reduced spectra were 
then subject to further
processing in ISAP\footnote{The ISO Spectral Analysis Package (ISAP) is 
a joint development by the LWS and SWS Instrument Teams and Data
Centers. Contributing institutes are CESR, IAS, IPAC, MPE, RAL and
SRON. ISAP is available from 
http://www.ipac.caltech.edu/iso/isap/isap.html.} \citep{stu98}
to remove further outliers and to perform flat-fielding and fringe
removal (where necessary) to produce the final spectral line
profiles from which line fluxes were measured.  The reduced data 
will also be made publicly available through the 
\iso\ Data Archive.

\begin{figure*}
	\includegraphics[width=5.5cm]{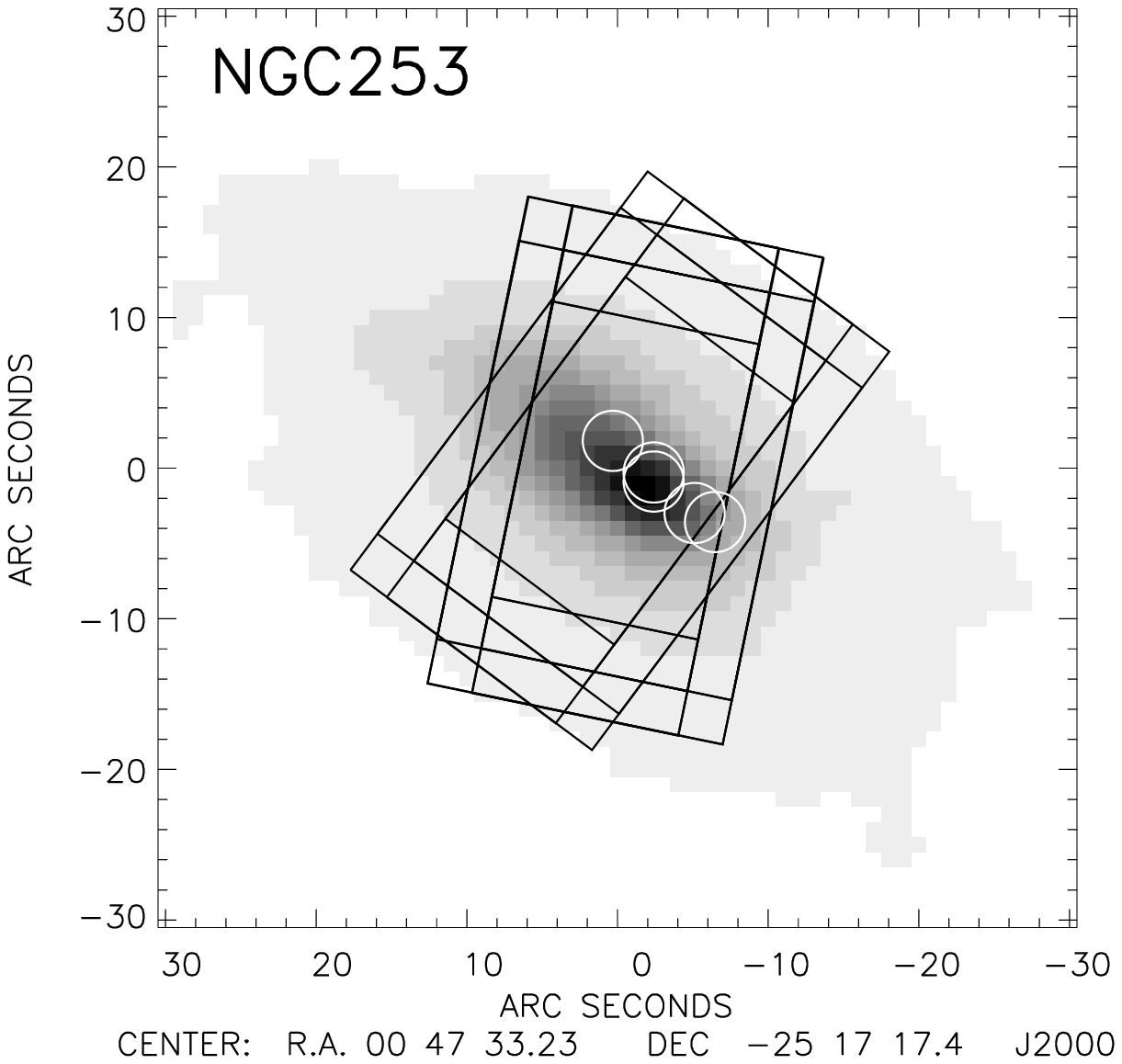}
	\includegraphics[width=5.5cm]{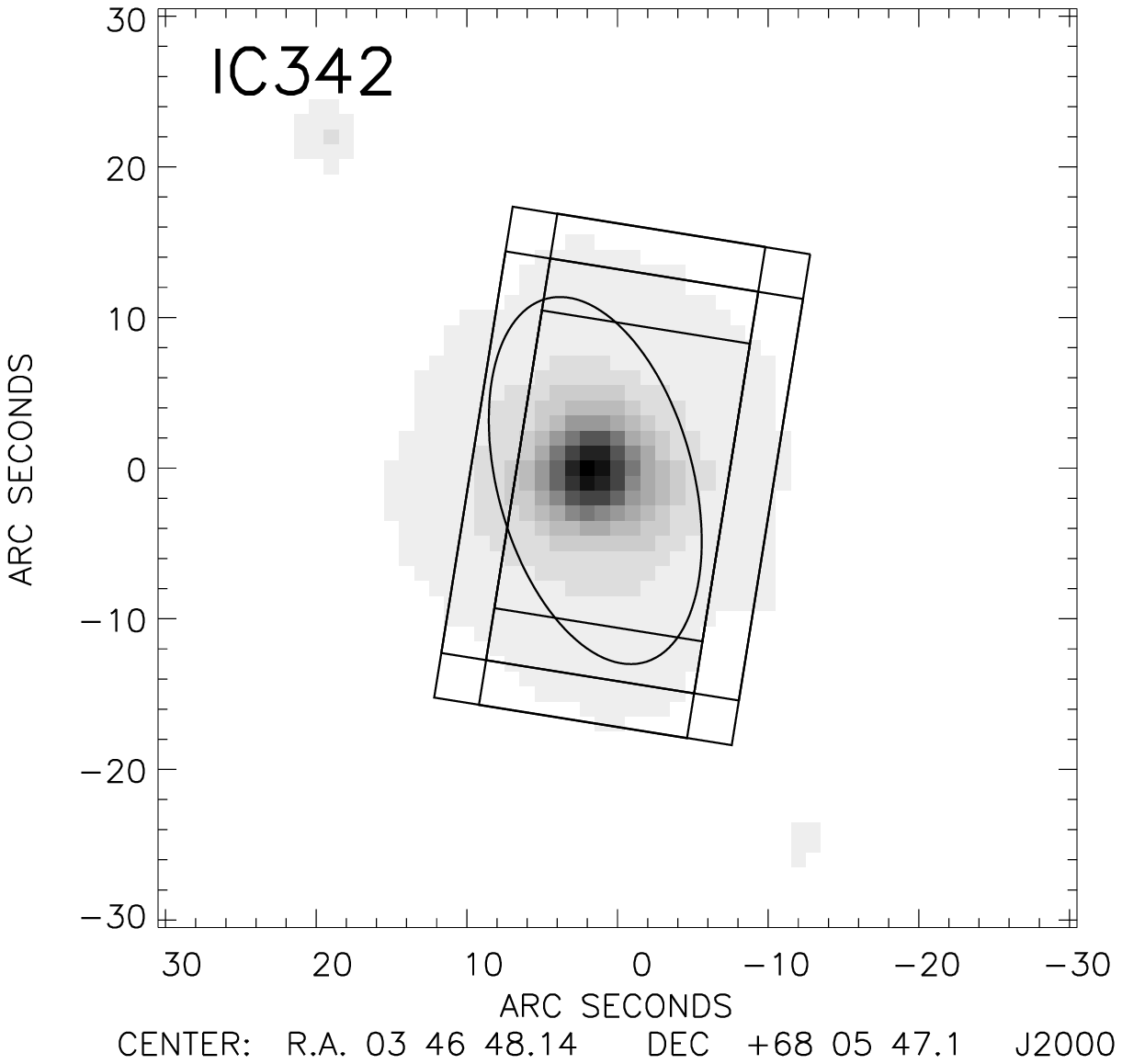}
	\includegraphics[width=5.5cm]{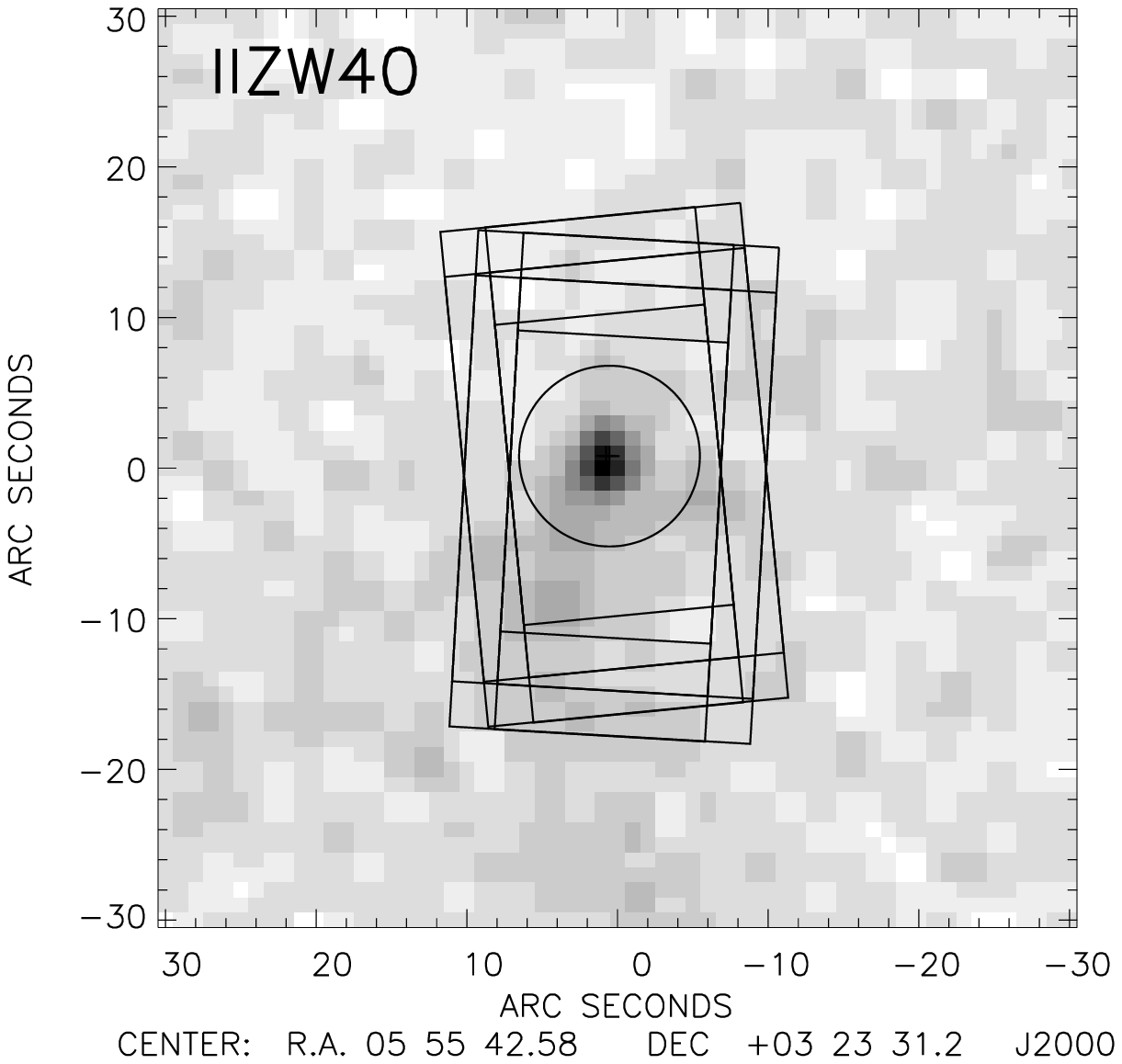}
	\includegraphics[width=5.5cm]{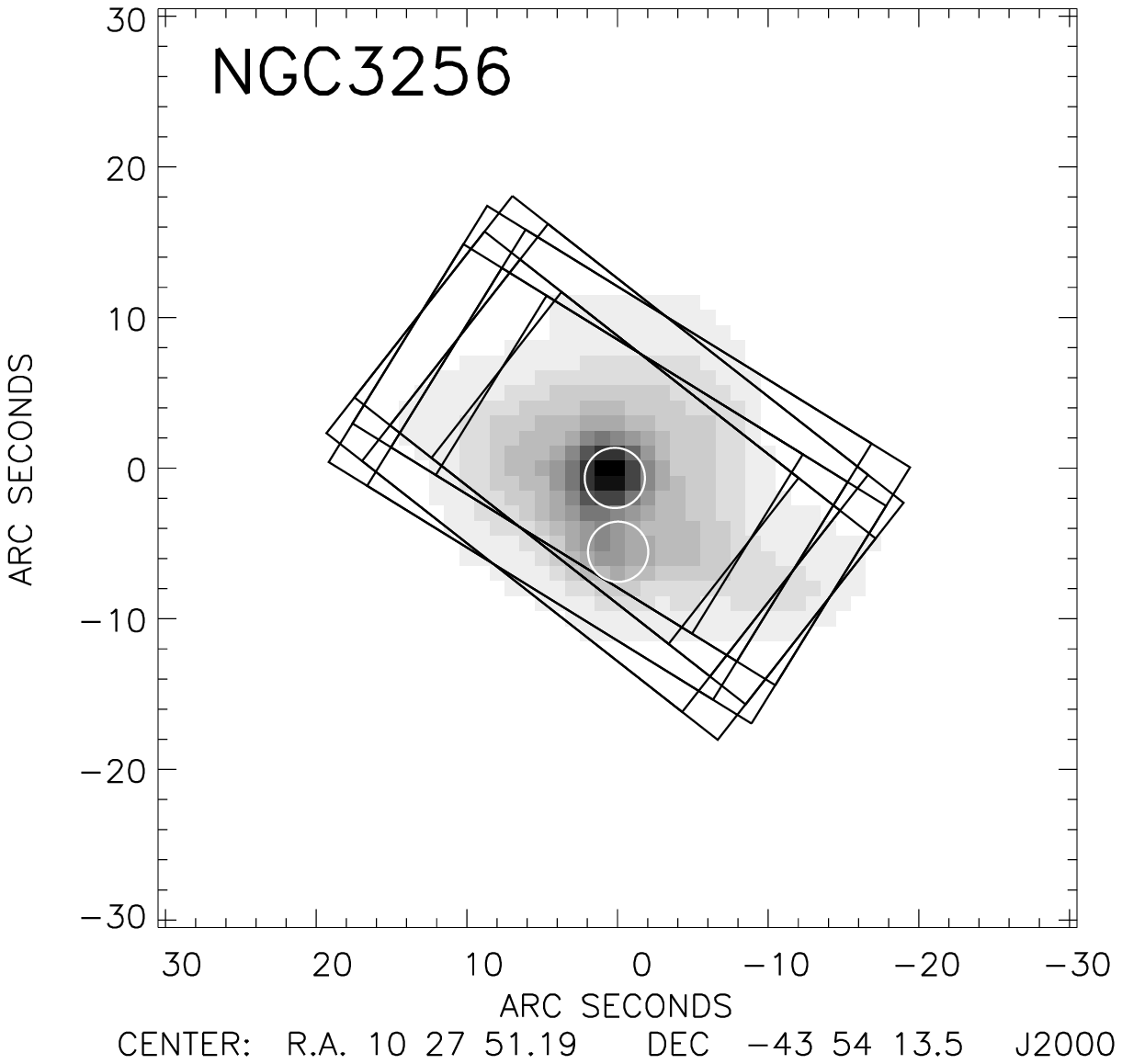}
	\includegraphics[width=5.5cm]{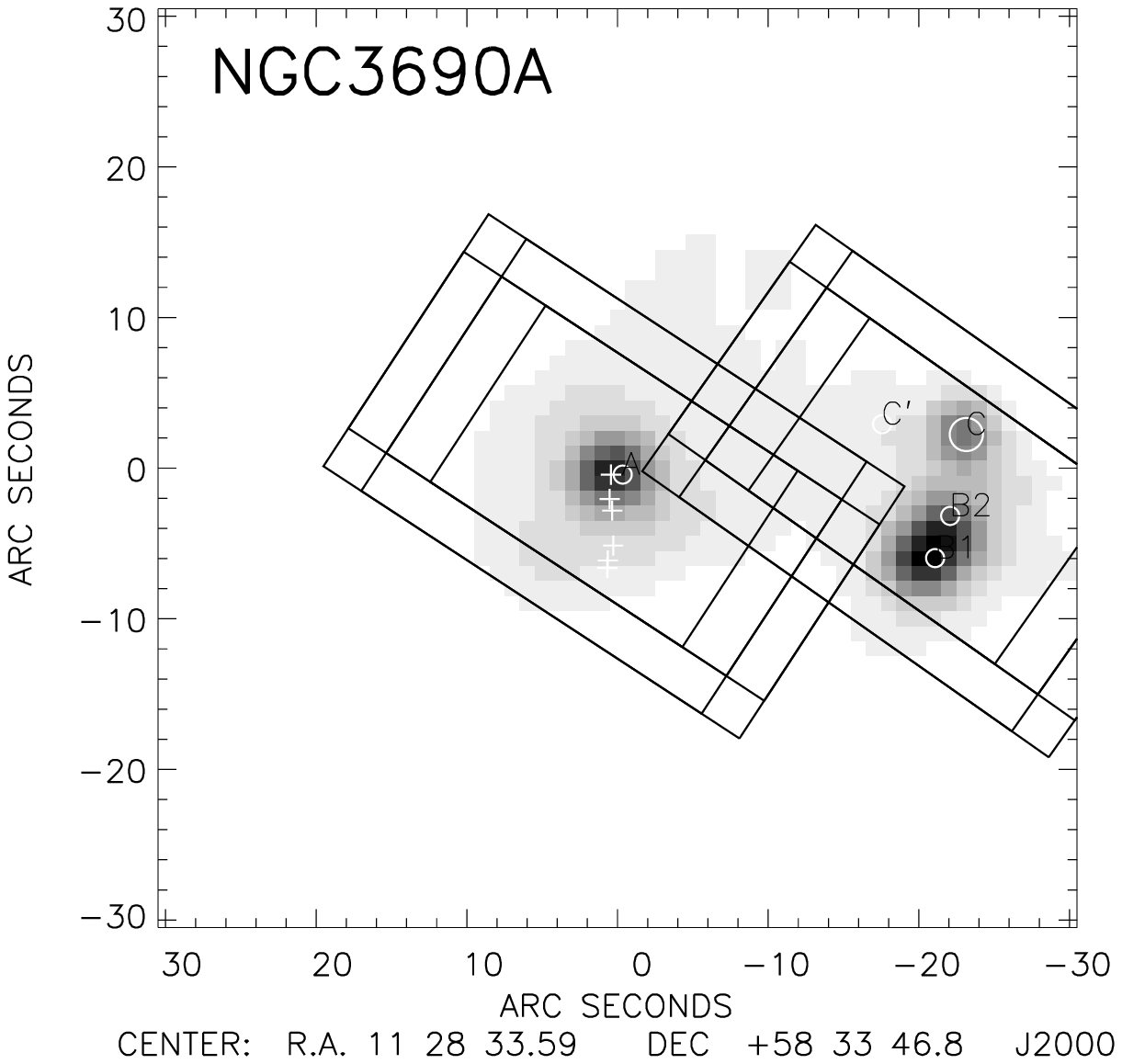}
	\includegraphics[width=5.5cm]{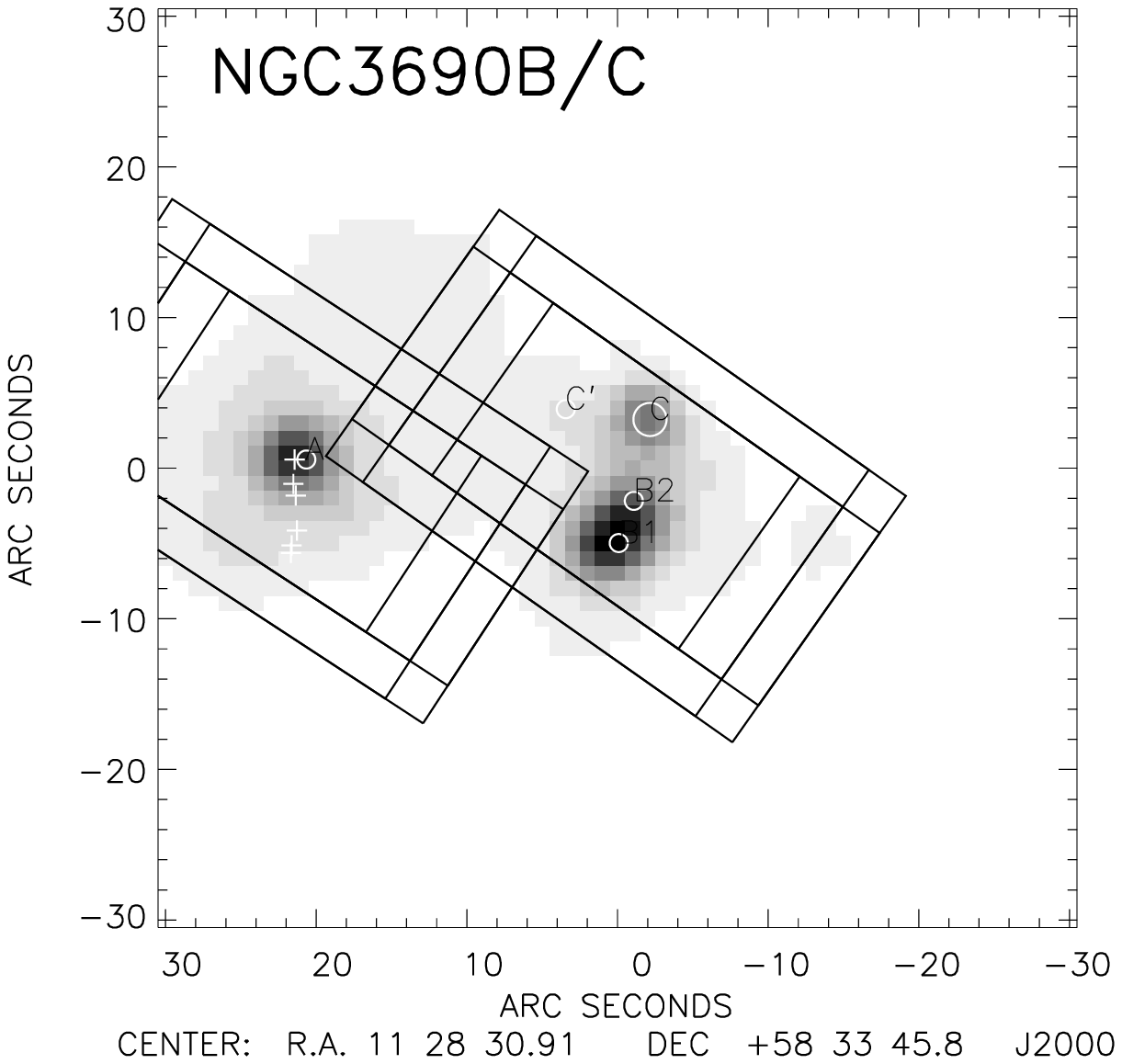}
	\includegraphics[width=5.5cm]{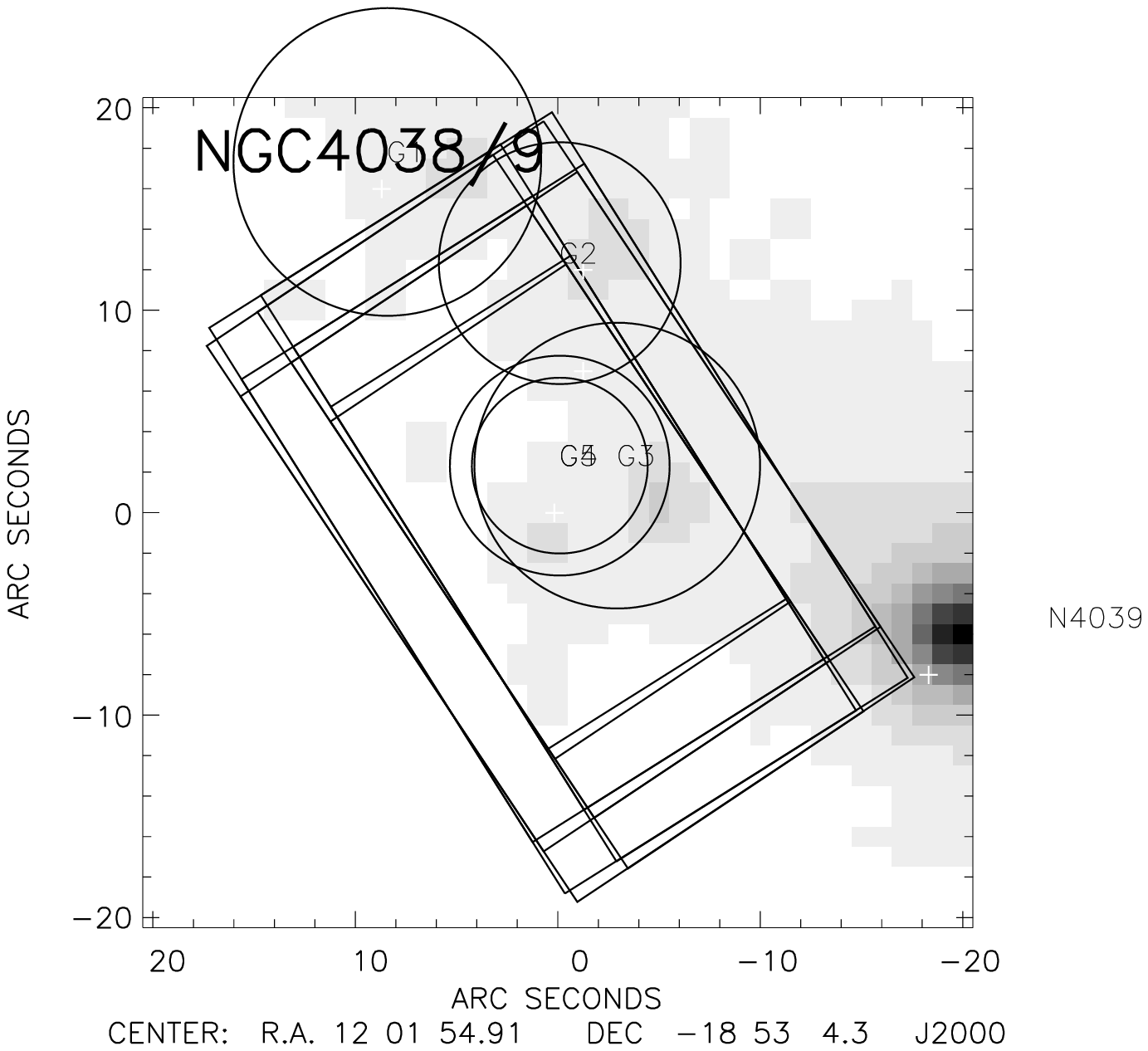}
	\includegraphics[width=5.5cm]{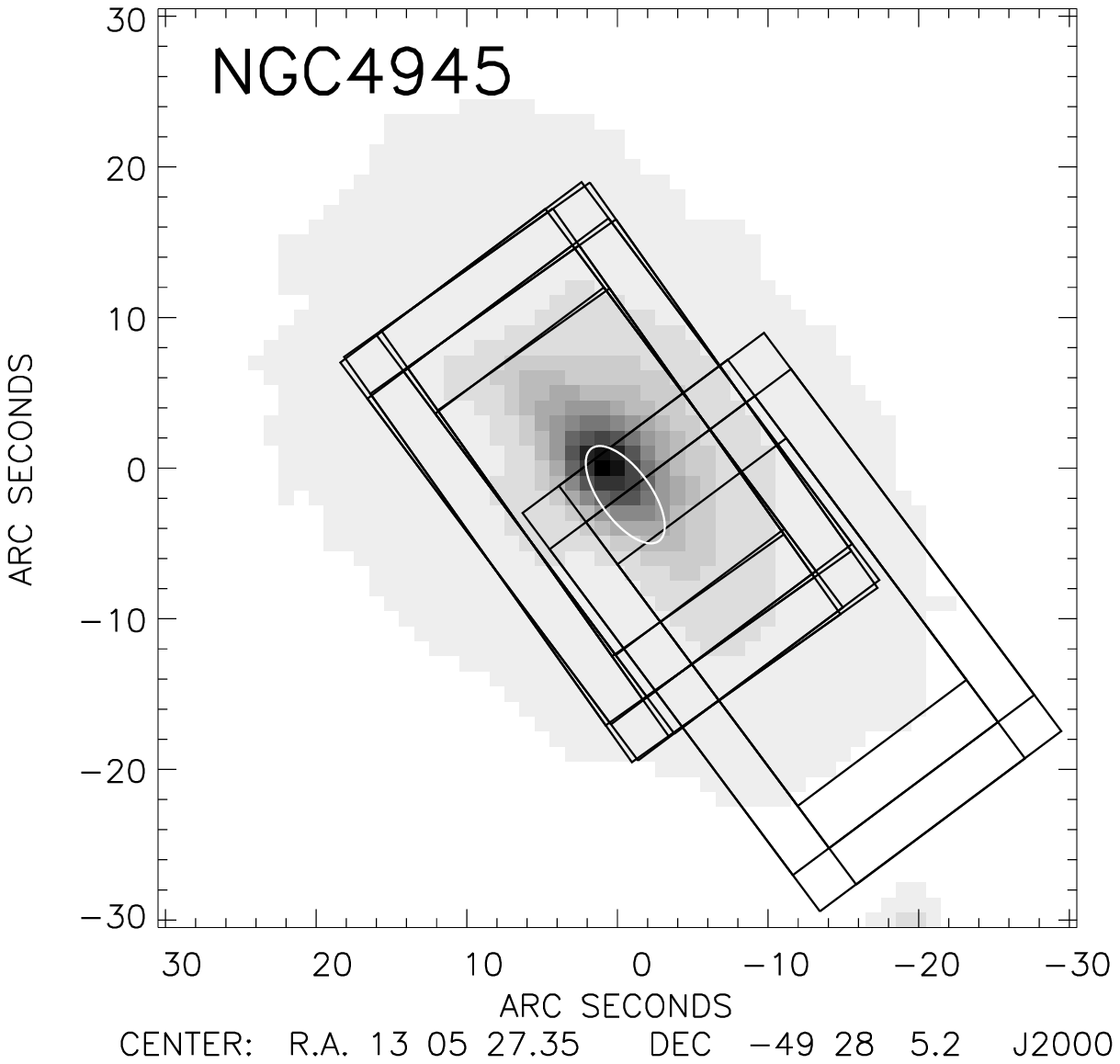}
	\includegraphics[width=5.5cm]{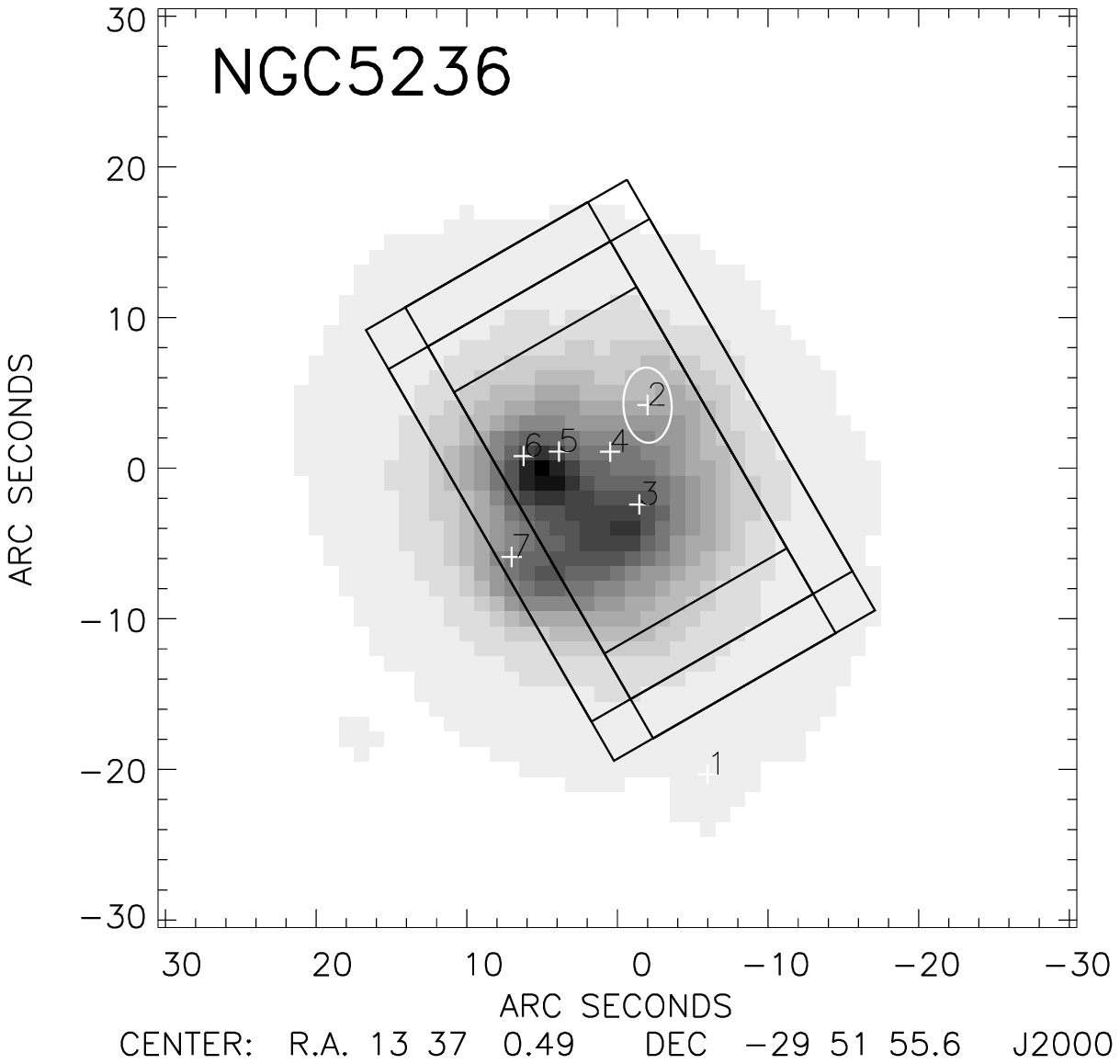}
	\includegraphics[width=5.5cm]{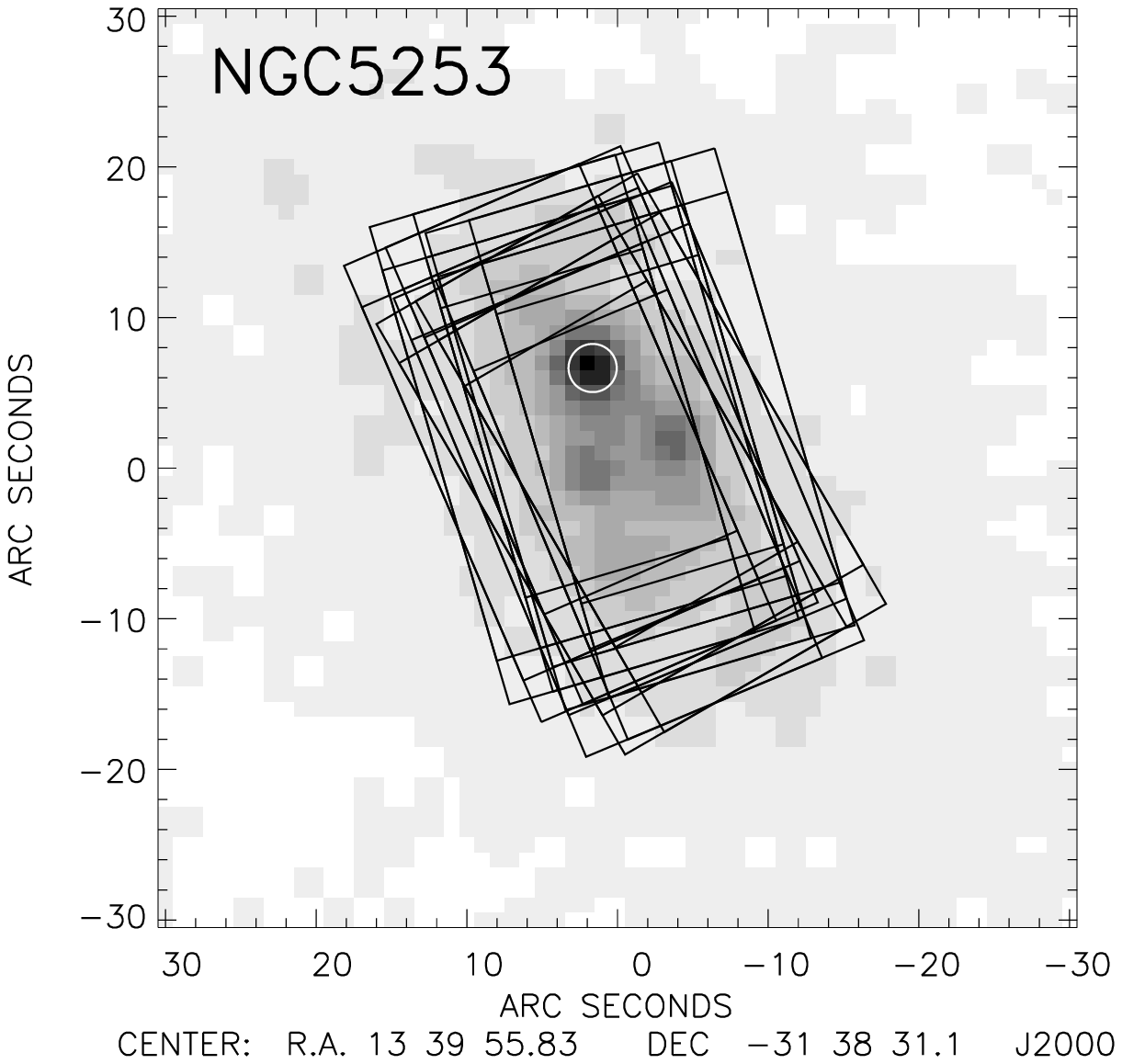}
	\includegraphics[width=5.5cm]{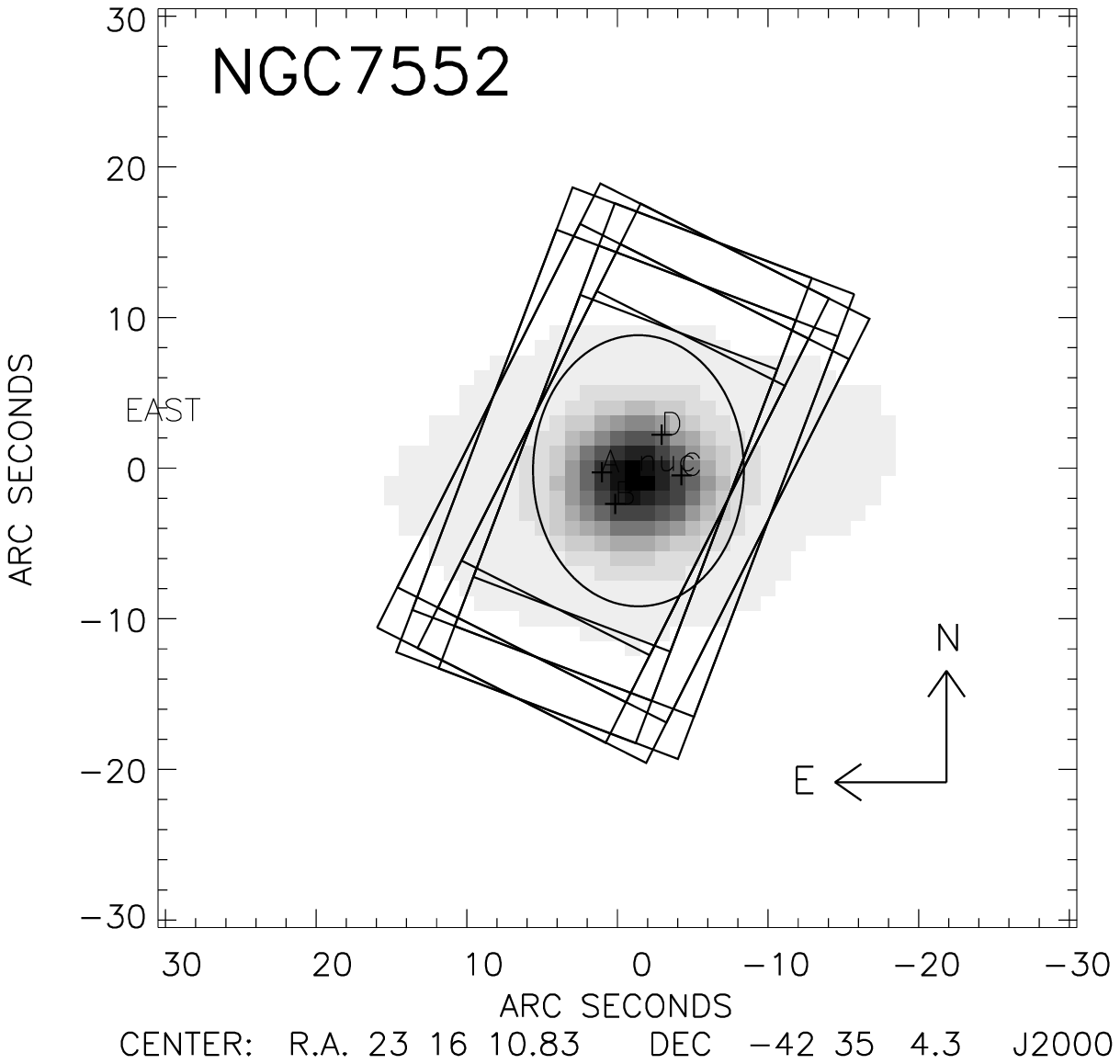}
	\caption{2MASS quick-look postage stamps (log-scale) of the starbursts 
with SWS apertures (rectangular) overlaid. Additionally, overlaid
ellipses indicate the extent of the starburst regions as obtained from
the literature.  Notes on individual sources:
{\em NGC 253} The MIR emission is dominated by a region of $\sim$10\arcsec\
in size \citep{bok98} and a single super-star cluster 
lying within it \citep{ket99}. 
{\em IC 342} The central radio source is of comparable size 
to the SWS aperture \citep{hum90}.
{\em II Zw 40} The  
12$\mu m$ emission originates from a
region only 0.5\arcsec\ in diameter \citep{beck02}. 
{\em M82} See \citet{nfs01}. 
{\em NGC 3256} Has a compact double nuclei with combined angular size
smaller than the SWS aperture \citep{nor95} and is in agreement with
the size ($<$10\arcsec\ diameter) determined from MIR mapping at
[NeII]12.8$\mu m$ and the N-band \citep{bok97}. 
{\em NGC 3690A \& B/C} The discrete compact radio ad CO emitting
sources are also the sites of strongest MIR emission - refer to
images presented in \citet{ket97} and \citet{soi01}. The overlaid
ellipses are the MIR source sizes determined by the latter authors.
{\em NGC 4038/9} The SWS aperture contains emission from SGMC2,3,4,5
as identified by \citet{wil00}.
{\em NGC 4945} N-band imaging reveals this source has a compact
unresolved nucleus with surrounding core emission oriented along the
galaxy's major axis \citep{kra01}. The overlaid ellipse marks the 
extent of the gas emitting CO(2-1)
lines from \citet{ott01}. Data from the SWS aperture offset
from the radio source are not included within the
analysis in this paper.
{\em NGC 5236} The ellipse marks the starburst extent determined from radio
maps from \citet{tur94} (additional hotspots are also marked).
{\em NGC 5253} extent of [SIV]10.5\micron\ emission from \citet{beck96}
\citep[but see also ][]{gor01}. 
{\em NGC 7552} The majority of the NIR and MIR flux originates from a
circumnuclear starburst ring of $\sim$10\arcsec\ in diameter \citep[ring
and hot spots are marked][]{for94,schin97}.
\label{aper}}
\end{figure*}

Since the \sws\ instrument takes measurements in a number of aperture
size and detector combinations, aperture corrections should be
considered for sources which are extended with respect to the
apertures (of sizes $14" \times 20"$, $14" \times 27"$ and
$20" \times 33"$ centred on the coordinates given in Table \ref{tabsam}).  
As we are interested in emission from only the compact nuclear 
optically obscured infrared starburst we used radio, millimetre or
\ir\ measurements 
from the literature to
determine the starburst extent to compare to the
\sws\ aperture (see Fig. \ref{aper}\footnote{This publication makes use of data products from the Two Micron All Sky Survey, which is a joint project of the University of Massachusetts and the
Infrared Processing and Analysis Center/California Institute of Technology, funded by the National Aeronautics and Space Administration and the
National Science Foundation} for details).
 We found no sources required correction 
for extension except M82. 
The corrections for
this source are
described in \citet{nfs01}.

Error propagation through the reduction of \sws\ data is
described in \citet{shi01} and \citet{lee02}.  To this error we add 
an additional 20\%
calibration error to account for systematic inaccuracies in the photometric
calibration \citep[Table 4.7 in][]{lee02}.

\section{Results - The MIR spectroscopic survey of local starbursts}
\label{survey}

We present a survey of mid-infrared emission lines from a sample of 
infrared starbursts.
The emission line fluxes for the most
commonly detected lines within the range of the \sws\ instrument
may be found in Table \ref{fluxtab}. 

Our starburst sample encompasses a variety of morphological types (Table
\ref{tabsam}) and includes six starbursts
exhibiting \wr\ spectral features \citep{scha99}. 
From this list of \wr\ galaxies we exclude \n5236 since \citet{ros86}
detected \wr\
signatures only in regions centred upon three supernovae sites, 
none of which lie within our observation aperture.
Recently,
\citet{bre02} report on the possible detection of \wr-like features in the 
spectrum of a nuclear `hot-spot' in \n5236.  However, due to the weakness 
of the features, for the subsequent analysis, we assume the
central region does not contain large numbers of \wr\ stars (see also 
Sect. \ref{wrsect}). 

Of our sources, only \n4945 displays evidence for harbouring an AGN.  
\citet{spo00} have demonstrated that
this AGN does not contribute to the flux of the \mir\
emission lines.  In addition, an AGN has been
advocated to be responsible for the ionisation required to
produce radio-recombination lines detected in \n253 \citep{moh02}.

We detect a number of \fsl s within the \sws\
spectra. In particular, the fine structure transitions of singly and doubly
ionised Ar and Ne are detected in the majority of
sources as well as those of \siii\ and \siv.  Other commonly detected
transitions are [O\,IV]25.89\mn, [Fe\,II]26.0\mn\ and [Si\,II]34.81\mn. 
The majority of these lines are likely to have an origin in electron
impact excitations in \hii\ regions that have been photo-ionised by
hot stars (assuming these sources are
pure starbursts) with a possible contribution from fast ionising
shocks. The [Si\,II] line is believed to partly originate in
photodissociation regions \citep[e.g.][]{ste95} at the interfaces
between H\,II regions and the
surrounding dense molecular clouds. The \sws\ detected lines span
a range of excitation potentials from 8.2eV for [Si\,II] to 55eV for [O\,IV].
The origin of the [O\,IV] transition in starburst and \wr\
galaxies is 
discussed in \citet{lut98} and \citet{scst99} respectively.
In addition, the \hrl s 
Brackett $\alpha$, $\beta$ and Pfund $\alpha$ are detected 
in most of our galaxies.  

We have also measured
several transitions of molecular hydrogen. $H_{2}$ data for all the
objects in our sample are included in the analysis of \citet{rig02}. 
Our reductions use the newest calibration files of the \sws\
instrument and we find the majority of our fluxes are consistent within the
calibration uncertainties of \citeauthor{rig02} and therefore we do not present
the $H_2$ line fluxes here. We also do not address the broad emission and
absorption features due to dust and ice that are measured in M\,82 and 
NGC\,253. These are discussed by
\citet{stu00}.

\begin{landscape}

\begin{table}
\renewcommand{\tabcolsep}{1.5mm}    

	\caption{Emission line fluxes (uncorrected for extinction), in units
of $10^{-20}Wcm^{-2}$. Also given are the calculated $A_{V}$ for a mixed
model with a \citet{lut99} extinction law and
$log_{10}<Q(H_{0})>$ calculated as described in the text.
\label{fluxtab}}

\begin{minipage}{25cm}

\vspace{3mm}

\begin{tabular}{lrrrrrrrrrrrrr}
\hline\hline

Line  & $\lambda$(\mn)  &  NGC253\footnote{Additional Lines ($\times
10^{-20}Wcm^{-2}$):[FeII]5.34017\mn\ 32.4, [PIII]17.8850\mn\ 3.0, [FeII]17.9359\mn\ 9.4} &  IC342 &  II Zw 40 &  M82\footnote{Additional Lines ($\times
10^{-20}Wcm^{-2}$):[FeII]5.34017\mn\ 43.2, [PIII]17.8850\mn\ 18.0,
[FeII]17.9359\mn\ 5.9, [FeIII]33.0384\mn\ $<20.$}&  NGC3256\footnote{Additional Lines ($\times
10^{-20}Wcm^{-2}$):[FeII]5.34017\mn\ 5.0, [PIII]17.8850\mn\ $<0.5$, [FeII]17.9359\mn\ 0.7} & 
 NGC3690A &  NGC3690B/C &  NGC4038 &  NGC4945 &  NGC5236 &  NGC5253 &  NGC7552\\
\hline

 HI Brb & 2.6260 &    14.7 &     3.4 &     1.2 &    41.0 &     4.2 &     2.1 &     2.1 &     0.8 &   --- &     4.8 &     2.7 &     2.5\\

 HI Pfd & 3.2970 &     2.6 &   --- &     $<$4.8 &     5.9 &   --- &   --- &   --- &   --- &   --- &   --- &   --- &   ---\\

 HI Pfg & 3.7410 &     3.0 &   --- &   --- &    10.7 &   --- &   --- &   --- &   --- &   --- &   --- &   --- &   ---\\

 HI Bra & 4.0520 &    31.2 &     7.0 &     2.6 &    81.5 &     4.4 &     2.6 &     3.7 &     2.5 &     9.9 &     7.8 &     4.5 &     4.4\\

 HI Pfa & 7.4600 &     7.6 &     2.4 &      $<$4.6 &    25.9 &     2.3 &     1.0 &     0.7 &   --- &     2.3 &   --- &     2.5 &   ---\\

 [MgVII] & 5.5032 &    $<$7.1 &     $<$0.8 &    $<$16.8 &    $<$69.3 &   --- &   --- &   --- &   --- &   --- &   --- &   --- &      $<$1.1\\

   [MgV] & 5.6098 &    $<$3.1 &   --- &    $<$17.4 &    $<$79.2 &     $<$2.3 &   --- &   --- &   --- &   --- &   --- &     $<$1.7 &   ---\\

  [ArII] & 6.9852 &   184.7 &    58.6 &     0.2 &   267.0 &    22.7 &
  --- &   --- &     2.2 &    {\em 24.2}\footnote{ This [ArII] flux is measured
from an SWS AOT01 spectrum which is not centred 
on the radio starburst, therefore Ar
fluxes and abundances for NGC 4945 are not used in the subsequent analysis.} &   --- &     2.3 &    18.5\\

  [NeVI] & 7.6524 &     $<$0.6 &   --- &     $<$6.2 &    $<$49.5 &     $<$1.3 &     $<$0.4 &     $<$0.5 &   --- &     $<$0.7 &   --- &   --- &   ---\\

 [ArIII] & 8.9913 &     7.0 &     2.6 &     2.0 &    47.6 &     4.7 &     1.3 &     2.8 &     1.8 &     $<$2.3 &   --- &     5.9 &     2.5\\

   [SIV] & 10.510 &     $<$3.3 &     $<$1.2 &    19.8 &    14.9 &     0.9 &     1.0 &     3.5 &     2.7 &     $<$0.9 &     0.9 &    34.0 &     0.3\\

  [NeII] & 12.813 &   453.6 &    90.7 &     1.7 &   714.0 &    89.2 &    31.9 &    27.7 &     7.7 &    92.5 &   133.9 &     7.9 &    68.0\\

   [NeV] & 14.321 &     $<$7.6 &     $<$0.7 &     $<$0.8 &    $<$29.7 &     $<$4.8 &     $<$0.5 &     $<$1.1 &   --- &     $<$0.5 &     $<$0.5 &     $<$0.3 &   ---\\

  [ClII] & 14.367 &     7.7 &     0.7 &     $<$4.1 &    $<$33.0 &     $<$1.8 &   --- &   --- &   --- &     0.3 &     1.8 &     $<$1.5 &   ---\\

 [NeIII] & 15.555 &    32.1 &     8.4 &    17.0 &   126.0 &    14.2 &     9.3 &    20.0 &     6.5 &     9.8 &     6.8 &    28.9 &     5.1\\

  [SIII] & 18.713 &    99.2 &    45.3 &     5.1 &   252.0 &    32.5 &     8.2 &    18.4 &     7.3 &     6.3 &    54.4 &    15.9 &    24.6\\

 [FeIII] & 22.925 &    17.1 &   --- &   --- &    46.6 &     3.3 &   --- &   --- &   --- &     0.7 &   --- &   --- &   ---\\

   [OIV] & 25.890 &     6.7 &     0.7 &     1.1 &     8.0 &     0.8 &     0.6 &     1.0 &     $<$0.3 &     3.5 &     0.8 &     0.9 &     1.0\\

  [FeII] & 25.988 &    14.2 &     4.8 &     1.0 &    36.7 &     2.8 &     1.6 &     2.2 &     0.3 &     3.5 &     4.2 &     0.9 &     2.0\\

  [SIII] & 33.481 &   197.3 &    89.3 &    11.4 &   562.0 &    56.4 &    27.1 &    34.1 &    16.0 &    51.4 &   109.8 &    34.4 &    41.1\\

  [SiII] & 34.815 &   273.2 &   131.6 &     4.5 &   839.0 &    84.0 &    43.4 &    42.2 &    10.5 &   107.4 &   138.0 &    16.7 &    61.5\\

 [NeIII] & 36.013 &     $<$4.5 &     $<$2.3 &     1.4 &    18.8 &     2.4 &     $<$1.3 &     1.8 &     1.8 &     $<$1.6 &     1.0 &     4.9 &     $<$0.8\\
\hline

 $A_{v}$ &         &  8.67 &   7.22 &  10.36 &  52.00\footnote{from
\citet{nfs01}}  &   $<$8.73 &  38.94 &  $<$57.13 &  37.34 & 188.19 &   5.00\footnote{from \citet{gen98}}  &   5.08 &  $<$73.90\\

   $log_{10}<Q(H_{0})>$  &         & 53.02 &  51.94 &  53.77 &  53.83 &  54.53 &  54.62 &  54.76 &  53.81 &  53.61 &  52.90 &  52.34 &  54.40\\

\hline

\end{tabular}
\end{minipage}
\end{table}

\end{landscape}

\section{Analysis}

\subsection{Extinction Correction}
\label{extcorr}

Several of our starburst galaxies are known to be dusty systems with large
obscuration; therefore the observed
line fluxes must be corrected for extinction.
To determine the
correction, we used the ratios of
$Br\beta/Br\alpha$ and $[SIII]18.7/33.5$. The intrinsic 
emissivities of \hrl s are 
theoretically well determined and may be compared
to the \sws\ measurements to determine the level of extinction.
Moreover, ratios of \hrl\ emissivities are relatively
insensitive to variations in electron temperature and density and 
therefore reasonable (rather
than precisely accurate) estimates of the two quantities is sufficient
to identify the appropriate intrinsic line ratio. Emissivity 
coefficients from \citet{sto95} for case B recombination with an 
electron temperature of $10^4$K
and density of $100cm^{-3}$ were used to calculate the intrinsic
line ratios for the H recombination lines.  
We chose not to use the \hrl\ ratio of 
$Br\alpha/Pf\alpha$ since \citet{lut99} suggest that the
extinction at 7.5$\mu$m may be higher than predicted by standard
extinction curves \citep[][]{dra89},
with the possibility of significant variation among sources.

The ratio $[SIII]18.7/33.5$ is more sensitive to density
than the H recombination lines but there are good reasons to assume
that the starbursts are close to the low density limit (Sect. \ref{ed}).  
We solved the rate equations for this ratio using the updated atomic data of
\citet{tay99} and confirmed that the resulting ratio is relatively 
insensitive to temperature. We adopt a value of $0.5$ for densities
close to the low density limit.

The extinction was calculated based upon a model in which obscuring material
and emitting gas are uniformly mixed. Such a `mixed' model is
considered more appropriate for a starburst region observed 
with our large apertures than a simple
foreground screen \citep[][]{mcl93}. The attenuation of the emitted 
intensity may be expressed as

\begin{equation}
\label{eq2}
I=I_{0}\frac{1-e^{-\tau_{\lambda}}}{\tau_{\lambda}}
\end{equation}                                                                 

where $\tau_{\lambda}$ is the optical depth at the wavelength of
interest ($\lambda$)
which is proportional to the optical depth in the V band $\tau_{V}$ through
the extinction law.
We assumed the Galactic centre extinction law of 
\citet{lut99} between 3-8.5\micron\ and adopted
a \citet{dra89} extinction law at all other wavelengths, with
a normalisation of the silicate feature of $\tau_{9.6}/\tau_{V}=0.122$.  
When comparing two recombination lines, the observed ratio may be
expressed in terms of $\tau_V$ using the appropriate emissivity and
extinction values. 

In the case
that all lines of both ratios $Br\beta/Br\alpha$ and
$[SIII]18.7/33.5$ were detected, we quote a weighted average
of the calculated $A_V$. For sources where only one of these ratios 
was measured, the
corresponding $A_V$ was adopted. For NGC3256, NGC3690B/C and NGC7552
both measured ratios were higher than the 
intrinsic ratios, implying $A_{V}=0$ which is unrealistic for our
dusty starbursts. We attribute the greater than intrinsic ratios to the large
errors 
on our flux measurements
($\sim 20\%$). 
Therefore, as no meaningful values for the extinction could be found for these sources,
we replaced the measured ratio by the measured ratio 
minus its associated error to determine a one sigma limit on $A_V$.  
For these sources we apply an extinction correction using the one
sigma limit of $A_V$ on the data but indicate the location of the
uncorrected data ($A_V=0$) in the figures presented 
in the following sections.  
The estimated $A_{V}$ used 
to correct all line fluxes are given in Table \ref{fluxtab}.

\subsection{Nebular Conditions}

\subsubsection{Electron density}
\label{ed}
To determine the electron density of our sources we used the ratios of 
[SIII]18.7\mn/33.5\mn\ and [NeIII]15.6\mn/36\mn\ and compared them to 
theoretical values obtained from a
solution to the rate equations, obtaining relations of line ratios and
densities close to those shown by \citet{ale99}, partly on the basis
of earlier atomic data.  Assuming a one-zone ionisation model, we
found that all of 
our starbursts
lie in the low density limit (i.e. well below the
critical density) within the errors. Using only [SIII] both in the context of density indication and in 
extinction correction estimates (Sect. \ref{extcorr}) allows degenerate 
low extinction/low density and high extinction/high density solutions. 
Low densities found from [NeIII] and general consistency of the 
recombination-line based and [SIII]-based
extinction estimates assure however, that it is reasonable to adopt 
low density conditions for the entire sample.    
Our computations of collisional
excitation indicate $n_{e} \sim 10-600 cm^{-3}$ therefore
we assumed an average electron density of $300cm^{-3}$. 
We were then able to use the 
expressions in the following section to calculate the nebular 
abundances in the low density limit (within which collisional
de-excitation may be ignored).  

\subsubsection{Electron Temperature}
\label{et}

As with density, electron temperature affects emissivities and thus 
inferred abundances albeit with only modest dependences for fine
structure transitions and \hrl s.  Since abundances are calculated as
ratios of \fsl\ to \hrl\ emissivities the weak dependence on
electron temperature partly cancels out \citep[e.g.,][]{givsm02}. 
Accurate determinations of electron temperature can be obtained from
recombination lines at radio and millimetre
wavelengths. Measurements in the latter wavelength range are easier to
interpret than in the radio since stimulated emission effects (which are present in the
radio) are negligible and the millimetre line flux 
primarily originates in spontaneous emission. Therefore we sought electron
temperature determinations based upon millimetre observations from the
literature; for the starburst galaxies M82 and NGC 253, electron
temperatures of $T_{e} \sim 5000K\pm1000K$ have
been determined by \citet{pux89} and \citet{pux97}. 
An electron temperature estimate based upon radio 
recombination lines is also available for NGC\,5253
\citep{moh01}, as well as estimates derived from
optical spectroscopy for NGC 5253 and II Zw 40 \citep{cam86,vac92,wal93}. 
The results for these two objects agree
well with the radio giving electron temperatures $\sim 10000-12500K$. 

The known electron temperatures in our starburst regions span a factor $\sim$2
consistent with the well known correlation between electron temperature and metallicity
\citep[e.g.][ for a sample of on-average lower metallicity objects]{cam86}.
For the remaining starburst galaxies, no accurate radio- or millimetre-based 
temperature determinations are available.  We have chosen not to rely
on electron temperature estimates derived from optical
spectroscopy since these may not reflect electron
temperatures of our obscured \ir-radio starbursts which cannot be
probed by optical spectroscopy.
For this reason, and to perform inter-comparisons between the
sources without introducing a bias due to uncertain adopted electron
temperatures, we elected to use a single `representative'
temperature for the entire sample of galaxies. Since M82 and NGC 253
are {\em archetypal} starbursts we chose our `representative' electron
temperature to be $T_{e}^{R} = 5000K$.  However, in doing so we appreciate that
deviations from this `representative' temperature are present and will affect
absolute abundances, thus distorting the
trends derived in the inter-comparisons. We will indicate these effects
(which are largest for high $T_{e}$ objects) in the appropriate location. In particular for the BCD 
galaxies, the magnitude of the effect of using 
$T_{e}$ of $\sim 10000K$ on the abundances is shown in the relevant plots.

For consistency we have also adopted the same `representative' electron temperature
($T_{e}^{R} = 5000K$)
for 
the \hii\ regions from \citet{giv02}, which we use as a comparison sample to 
the starbursts. We have
re-calculated extinctions and abundances for this data set in order to
eliminate any discrepancies due to different assumptions.  The \hii\
regions and starbursts are treated differently only in the obscuring
model used to determine the extinction correction. For \hii\
regions, often located in the Galactic plane but far from the Sun, 
we have adopted a uniform foreground screen model since 
we assume that extinction due to the foreground galactic ISM exceeds the
extinction due to dust mixed within the ionised gas of the \hii\
region itself.
Similarly to
the BCDs, the assumed representative temperature (5000K) is below the values that have been 
previously used [10000K
\citep{giv02}, 7500K \citep{mar02}].  Therefore the effect of
increasing the temperature to 10000K shown on the figures in the next
section is also applicable to part of the \citet{giv02} sample of H\,II regions, 
namely those at large Galactocentric radii or inside the Magellanic
Clouds.

\subsubsection{Lyman continuum photon emission rate}
The Lyman continuum photon emission rate [$Q(H_{0})$] is the number of 
photons emerging per second
from a star cluster with sufficient energies to ionise hydrogen.
We have used H recombination lines to determine the average hydrogen
ionising rate [$Q(H_{0})$] using 

\begin{equation}
\label{eq4}
Q(H_0) \sim \frac{L_{line}}{h\nu_{line}} \frac{\alpha_{B}(T_{e})}{\alpha^{eff}_{line}(T_{e})}.
\end{equation}

The ratio of recombination
coefficients ($\alpha_{B}/\alpha^{eff}_{line}$) is relatively 
insensitive to temperature \citep[see
equation 5.23][]{ost89}.  The recombination coefficients were taken
from \citet{sto95}. 
The calculated values of $Q(H_0)$ are given in Table \ref{fluxtab}.

\subsection{Excitation}

\begin{figure}
	\resizebox{\hsize}{!}{\includegraphics{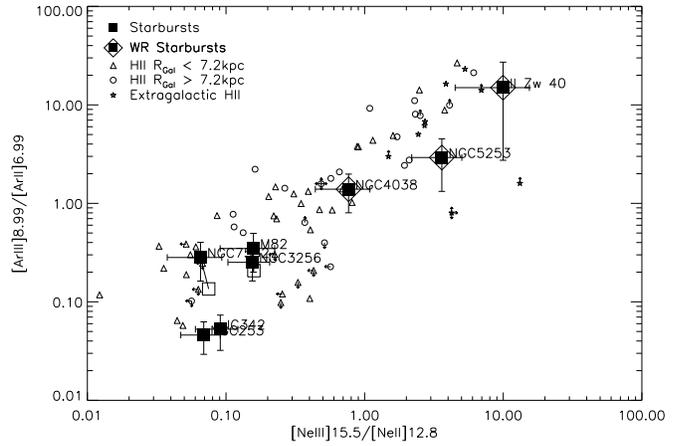}}
	\caption{Excitation diagnostic: $[NeIII]15.5/[NeII]12.8$
	versus $[ArIII]8.9/[ArII]6.9$. Starbursts are plotted as 
	filled squares
	with the \wr\ starbursts marked by an enclosing diamond.
	Galactic HII region data from
	\citet{giv02} are separated at the median galactocentric
	radius to provide `inner' (open triangles) and `outer'
	(open circles) samples. Finally, we plot the
	extragalactic H\,II regions from the same reference as
	open stars. The same key is used to distinguish the
	data points in all subsequent plots.
	Within this diagram low excitation is
	to the bottom-left quadrant and high excitation is to the
	top-right quadrant.  All fluxes involved in this diagram 
	were corrected for extinction. For NGC3256, NGC3690 B/C and
	NGC 7552 the effect of adopting $A_V=0$ on the
	derived quantities is plotted as an open square.
	\label{nearex}}
\end{figure}

\begin{figure}
	\resizebox{\hsize}{!}{\includegraphics{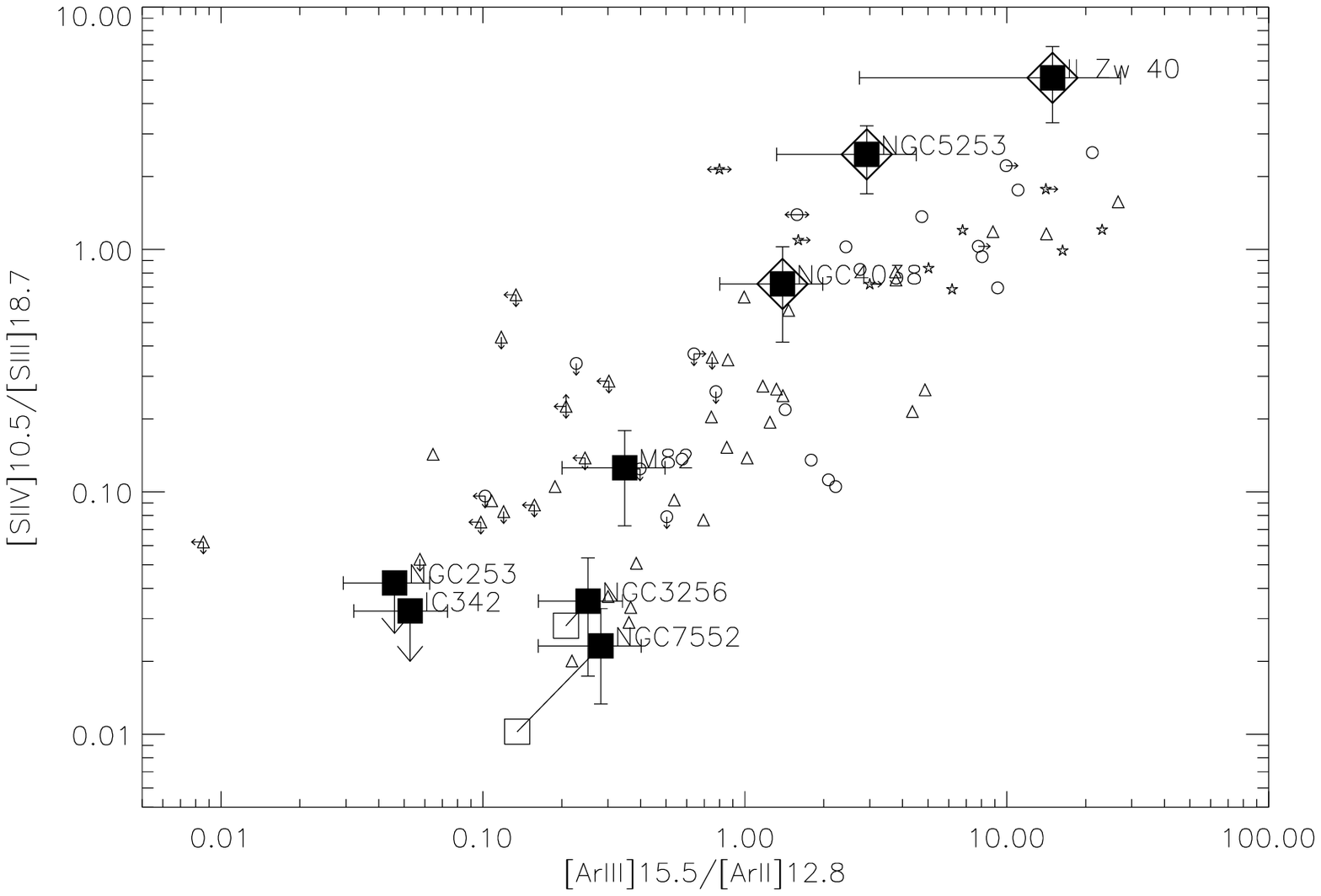}}
	\caption{Excitation diagnostic: $[ArIII]8.9/[ArII]6.9$ versus
	$[SIV]10.5/[SIII]18.7$. Key as in Figure \ref{nearex}. 
	Within this diagram low excitation is
	to the bottom-left quadrant and high excitation is to the
	top-right quadrant.  All fluxes involved in this diagram were corrected for extinction.
	\label{arsex}}
\end{figure}

\begin{figure}
	\resizebox{\hsize}{!}{\includegraphics{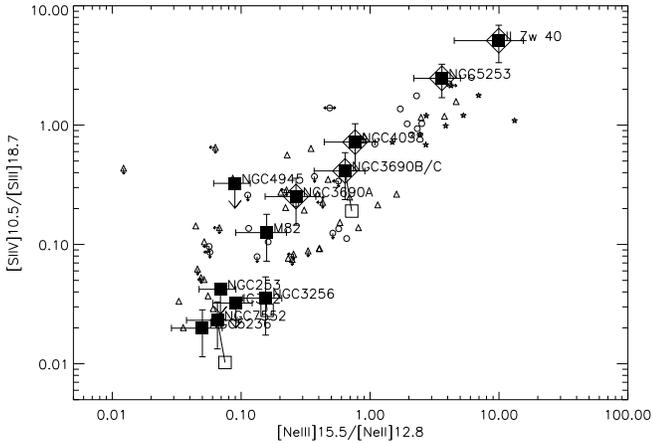}}
	\caption{Excitation diagnostic: $[NeIII]15.5/[NeII]12.8$
	versus $[SIV]10.5/[SIII]18.7$. Key as in Figure \ref{nearex}.
	Within this diagram low excitation is
	to the bottom-left quadrant and high excitation is to the
	top-right quadrant.  All fluxes involved in this diagram were corrected for extinction.
	\label{sneex}}
\end{figure}

We may investigate excitation within the starburst sample using 
ratios of \fsl\ fluxes of different ionic species of the same
element (e.g. $X^{+i+1}/X^{+i}$). This ratio, to the first order 
and for a given ionisation
parameter, is proportional to the number of photons which have
sufficient energy to ionise the state $X^{+i}$ relative to the
number of Lyman continuum photons. The
difference in ionisation potentials between the ionic species means 
that ratios of these lines are sensitive to 
the shape of the spectral energy distribution (SED) of the ionising source. 
We have used (and combined)
ratios of some of the strongest lines, $[NeIII]15.5/[NeII]12.8$, $[ArIII]8.9/[ArII]6.9$ and
$[SIV]10.5/[SIII]18.7$, from our \mir\ survey. 
Ar$^+$, S$^{++}$ 
and Ne$^+$ have ionisation
potentials corresponding to $\sim2$, $ \sim2.5$ and $\sim3$ 
Rydbergs, respectively. Therefore these line ratios can constrain the ionising spectra over a range of a
few Rydbergs.

We have formed diagnostic excitation planes by combining the \fsl\
ratios as shown in
Figs. \ref{nearex}, \ref{arsex} and \ref{sneex}.
The starbursts 
are relatively spread out across all three excitation planes.  The majority
of sources occupy the lower left corner of the planes, which indicates that
these have low excitation.  Extending to higher excitation in
both axes are the BCD galaxies and
those that exhibit \wr\ features (\wr\ galaxies
are discussed in more detail in Sect. \ref{wrsect}).  This result 
is consistent with existing
knowledge of both of these source types \citep{scha99}.

Also plotted in the excitation planes are extinction corrected data of
\hii\ regions from \citet{giv02}. The authors found an increasing excitation gradient with
galactocentric radius within the galaxy \citep[see
also][]{mar02,givsm02}. Therefore these data have 
been separated into two
broad bins based upon galactocentric radius (denoted by `inner' and
`outer') and reflect low and high
excitation.
The distribution of the starbursts is well matched
with the \hii\ region data. The locations of the `inner' H\,II regions 
is similar to those of low excitation starburst
galaxies.  The exceptions are the \wr\ and BCD galaxies which are 
located in regions similar to those of the `outer' and extragalactic H\,II 
regions. These extragalactic \hii\ regions have been
heterogeneously selected from the Magellanic Clouds \citep{giv02}. The 
SMC and LMC are both low metallicity dwarf galaxies containing
\wr\ stars and thus the coincidence of our \wr\ and BCD starbursts with  
the extragalactic \hii\ regions is as expected.

For both the starbursts and the \hii\ regions we see a general
correlation between the three excitation ratios of Ne, Ar and S in
Figs. \ref{nearex}-\ref{sneex}.  
The existence of a correlation between all three ratios implies that
they probe the same excitation mechanism. Moreover, the ratios are sensitive to photons emitted $\ga 2$,
$\ga 2.5$ and
$\ga 3$ Rydbergs and thus the correlations suggest that regions with ionisation
potentials between $2-3$ Rydbergs in the ionising spectra, are 
likely to be closely related in `hardness'.

\subsection{Elemental Abundances}

In the following, we derive the elemental abundances for our sample
galaxies and discuss their relation to other factors, in particular
the role of metallicity in determining the wide range of excitations. 
Although we detect \fsl s for several elements in some objects
(Table~\ref{fluxtab}), we 
restrict our abundance analysis to only those lines originating from \hii\
regions and detected in the majority of sources within our sample
i.e. neon, argon and sulphur, which   
are all thought to be `primary' products of
nucleosynthesis. They are products of oxygen burning, possibly in the
late stages of evolution of massive stars \citep{gar02}. Thus the
abundance of these primary products are likely to be correlated with
the abundance of oxygen and should trace metallicity well. 

\subsubsection{Prescription}
We calculate abundances using the following prescription. 
In general, the flux ($F^{X^{+i}}$) of a given ionic species
($X^{+i}$) may be written as the product of the number
density of the ionic species ($n^{X^{+i}}$), the electron density of the
ionised gas ($n_e$) and
$j^{X^{+i}}$ - the emissivity of the species, i.e.,

\begin{equation}
\frac{F_{\lambda}^{X^{+i}}}{F_{\lambda}^{H^+}} = \frac{ n^{X^{+i}} n_e  j_{\lambda}^{X^{+i}}}{ n^{H^+} n_e  j_{\lambda}^{H^{+}}}
\label{eq5}
\end{equation}

The ionic abundance ($n^{X^{+i}}/n^{H^+}$) may then be calculated from the 
flux ratio of the \fsl s of an ionic species to a \hrl ,
multiplied by the ratio of the emissivities.  In \hii\ regions nearly 
all the hydrogen is
ionised, therefore $n^{H+} \sim n^{H}$. The emissivity of the 
\hrl\ is obtained from the tables of \citet{sto95}. 
Strictly speaking, the emissivities of ionic \fsl s should be obtained 
by solving the rate equations for a full multilevel system under the
appropriate conditions. However, for 
ground state fine structure transitions and a low density \hii\ 
region, the following simple formula provides a good approximation

\begin{equation}
j_{\lambda}^{X^{+i}} = \sum_{u}\frac{hc}{\lambda}\left(\frac{8.629\times10^{-6}}{T_e^{1/2}}\right)^{1/2}\left(\frac{\Upsilon_{lu}}{\omega_{l}}\right)e^{-\frac{\chi_{lu}}{kT_e}}
\label{eq6}
\end{equation}

where $j$ has units of $erg\ cm^3\ s^{-1}$, $\Upsilon_{lu}$ is the
effective collision strength, $\omega_l$ is the statistical weight of 
the lower level, 
$\chi_{lu}$ is the energy difference to the collisionally excited
level from the ground state, and the summation is over 
those fine structure levels which, when collisionally excited from the ground
state, decay through the line of interest. 
For the lines we used (Table~\ref{cotab}), it is
sufficient to consider one upper level only, with the exception of the
[NeIII]15.5\mn\ and [ArIII]9.0\mn\ lines where two levels are needed. 
Effective collision strengths were obtained from the 
appropriate reference of the IRON project \citep[][ Table \ref{cotab}]{hum93}. 
The abundance of an ionic species may be calculated using

\begin{equation}
\frac{n^{X^{+i}}}{n^{H}} =\left[\frac{F_{\lambda}^{X^{+i}}}{F_{\lambda}^{H^+}}\right]\left[\frac{j_{\lambda}^{H^{+}}}{j_{\lambda}^{X^{+i}}}\right].
\label{eq7}
\end{equation}

Once an ionic abundance has been estimated using the
formula above (see Table \ref{abtab}), we sum the abundances for each
ion from a 
given element and apply the
ionisation correction factors (hereafter ICF) from \citet{mar02} to 
correct for the contribution from unobserved
ionic species.  The application of the ICFs from \citet{mar02}
is appropriate for these data since we are using the same ionic species
to determine the abundance, and the model grid used 
to calculate the ICFs is sufficient to include the sources similar to
ours (i.e. those in the low density limit).
Since the two dominant ionisation stages of Ne, Ar and S for \hii\
region 
conditions are  
observed in our galaxies, the ionisation
corrections factors are always close to unity. Following
\citet{mar02}, we do not correct our Ne abundances as the ICF is
$\sim 1$ \citep[see also][]{giv02} and for S we applied a factor of 1.15. The ICF for
Ar is seen to increase with excitation, from 1 to a maximum of 1.35. We
interpolate over the factors calculated by \citet[][ Fig. 12]{mar02}
to obtain an ICF appropriate for the excitation (as determined from
the Ne ratio) of each of
our sources.

\begin{table*}
	\caption{Atomic data used to calculate the fine structure line 
emissivities and abundances. Effective Collision Strengths are interpolated
to a temperature of 5000K. In addition the $Br\alpha$
recombination-line was used, adopting an emissivity of
$\epsilon_{\lambda}^{Br\alpha}=2.298\times10^{-26}ergcm^{3}s^{-1}$ from
\citet{sto95} for $n_e\sim100cm^-3$ and $T_{e}\sim5000K$. 
\label{cotab}}
\begin{tabular}{lrlrrl}
\hline\hline
Line     &Wavelength&$\Upsilon_{lu}$&$\omega_l$&$\chi_{lu}$&Reference for $\Upsilon_{ul}$\\
         &	(\mn)    &               &          &(eV)         &\\
\hline
{[NeII]} &  12.8136 & 0.277         &        4 &0.09684    &\citet{sar94}\\
{[NeIII]}&  15.5551 & 0.730         &        5 &0.07971    &\citet{but94}\\
         &          & 0.199         &        5 &0.11413    &\citet{but94}\\ 
{[ArII]} &   6.9853 & 2.70          &        4 &0.17749    &\citet{pel95}\\
{[ArIII]}&   8.9914 & 3.182         &        5 &0.13789    &\citet{galv95}\\
         &          & 0.663         &        5 &0.19469    &\citet{galv95}\\
{[SIII]} &  18.7130 & 1.41          &        1 &0.10329    &\citet{tay99}\\
{[SIV]}  &  10.5105 & 8.10          &        2 &0.11796    &\citet{sar99}\\
\hline
\end{tabular}
\end{table*}

\begin{table*}
\renewcommand{\tabcolsep}{1.5mm}    

	\caption{Calculated elemental abundances for each source, per 
ionic species and the total (ICF corrected) abundance with reference to 
solar abundances
([X/H]): for Ne, Ar and S from \citet{gre98} (first line). The abundances
presented are for $T_{e}=5000K$. See text for the  
effect of this assumption on the lowest abundances.
\label{abtab}}
\begin{tabular}{lrrrrrrrrr}
\hline
Name & $n^{Ne^{+}}/n^{H}$ & $n^{Ne^{++}}/n^{H}$ & [Ne/H] & $n^{Ar^{+}}/n^{H}$
 & $n^{Ar^{++}}/n^{H}$ & [Ar/H] & $n^{S^{++}}/n^{H}$ & $n^{S^{3+}}/n^{H}$ & 
[S/H]\\

 & $\times 10^{-5}$ &  $\times 10^{-6}$ & &   $\times 10^{-6}$    &
$\times 10^{-7}$ & &  $\times 10{-6}$    &
$\times 10^{-7}$ & $\times 10^{-1}$ \\
\hline
             Solar &  &  & 1.20e-04 &  &  & 2.51e-06 &  &  & 1.58e-05\\
\hline
             IC342 & $    27.6 \pm      8.7$ & $    10.9 \pm      3.8$ & $     2.4 \pm      0.7$ & $    12.1 \pm      3.8$ & $     6.6 \pm      2.6$ & $     5.1 \pm      1.5$ & $     9.4 \pm      3.0$ & $     ----$ & $     6.8 \pm      2.1$  \\
          II Zw 40 & $     1.4 \pm      0.8$ & $    59.0 \pm     20.8$ & $     0.6 \pm      0.2$ & $     0.1 \pm      0.1$ & $    14.6 \pm      6.0$ & $     0.9 \pm      0.3$ & $     2.9 \pm      1.1$ & $    32.3 \pm     10.8$ & $     4.4 \pm      1.1$  \\
               M82 & $    17.0 \pm      7.2$ & $    11.6 \pm      4.9$ & $     1.5 \pm      0.6$ & $     4.4 \pm      1.9$ & $    15.8 \pm      6.7$ & $     2.4 \pm      0.8$ & $     4.3 \pm      1.8$ & $     1.2 \pm      0.5$ & $     3.2 \pm      1.3$  \\
            NGC253 & $    31.0 \pm      9.4$ & $     9.3 \pm      3.1$ & $     2.7 \pm      0.8$ & $     8.6 \pm      2.7$ & $     4.1 \pm      1.5$ & $     3.6 \pm      1.1$ & $     4.6 \pm      1.5$ & $   ----$ & $     3.4 \pm      1.1$  \\
           NGC3256 & $    43.2 \pm     18.5$ & $    29.1 \pm     13.1$ & $     3.8 \pm      1.5$ & $     7.4 \pm      3.3$ & $    19.3 \pm      8.9$ & $     3.8 \pm      1.4$ & $    10.7 \pm      4.7$ & $     0.8 \pm      0.5$ & $     7.8 \pm      3.4$  \\
          NGC3690A & $    23.8 \pm     10.1$ & $    27.6 \pm     11.7$ & $     2.2 \pm      0.8$ & $   ----$ & $    12.9 \pm      5.5$ & $     0.5 \pm      0.2$ & $     4.4 \pm      1.9$ & $     2.4 \pm      1.0$ & $     3.4 \pm      1.4$  \\
        NGC3690B/C & $    14.4 \pm      6.1$ & $    40.2 \pm     17.0$ & $     1.5 \pm      0.5$ & $ ----$ & $    21.1 \pm      9.0$ & $     1.0 \pm      0.4$ & $     7.0 \pm      3.0$ & $     6.4 \pm      2.7$ & $     5.5 \pm      2.2$  \\
           NGC4038 & $     6.1 \pm      2.6$ & $    20.3 \pm      8.6$ & $     0.7 \pm      0.2$ & $     1.2 \pm      0.5$ & $    17.6 \pm      7.5$ & $     1.4 \pm      0.4$ & $     4.2 \pm      1.8$ & $     6.6 \pm      2.8$ & $     3.5 \pm      1.3$  \\
           NGC4945 & $    17.2 \pm      6.5$ & $     6.7 \pm      2.6$
& $     1.5 \pm      0.5$ & $     ----$ & $  ----$ & 
$----$ & $     0.9 \pm      0.4$ & $   ----$ & $     0.6 \pm      0.3$  \\

           NGC5236 & $    36.8 \pm     15.6$ & $     8.0 \pm      3.4$ & $     3.1 \pm      1.3$ & $     ----$ & $    ----$ & $   ----$ & $    10.1 \pm      4.3$ & $     0.4 \pm      0.2$ & $     7.4 \pm      3.1$  \\
           NGC5253 & $     3.8 \pm      1.7$ & $    59.4 \pm     22.3$ & $     0.8 \pm      0.2$ & $     0.7 \pm      0.4$ & $    22.2 \pm      9.0$ & $     1.5 \pm      0.5$ & $     5.2 \pm      2.0$ & $    28.1 \pm     10.5$ & $     5.8 \pm      1.6$  \\
           NGC7552 & $    29.5 \pm     12.5$ & $     8.4 \pm      3.6$ & $     2.5 \pm      1.0$ & $     5.6 \pm      2.4$ & $    16.3 \pm      6.9$ & $     2.9 \pm      1.0$ & $     7.9 \pm      3.3$ & $     0.4 \pm      0.2$ & $     5.7 \pm      2.4$  \\

\hline
\end{tabular}
\end{table*}

\subsubsection{The abundance of neon and argon}
\label{near}

\begin{figure}
	\resizebox{\hsize}{!}{\includegraphics{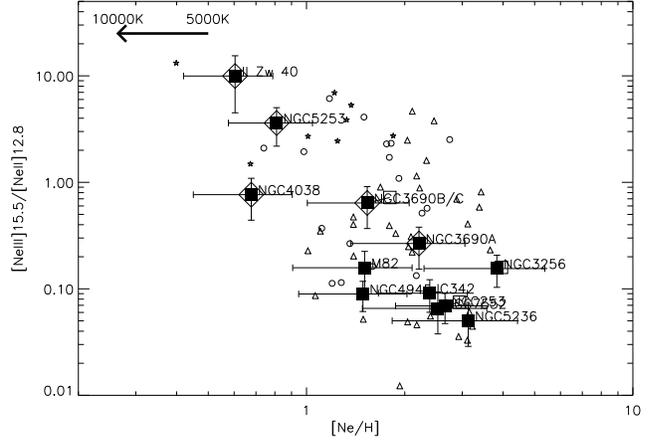}}
	\caption{Excitation against abundance for
	neon. Starbursts are plotted as asterisks
	with the \wr\ starbursts marked by an enclosing diamond.
	Galactic and local-extragalactic HII region data from
	\citet{giv02} are also plotted as open triangles, circles and
	stars. For a given
	metallicity the starbursts are of lower excitation than the
	HII regions. The arrow indicates the effect on the abundances of 
        changing the 
        adopted electron temperature from 5000 to 10000\,K.   
	\label{neab}}
\end{figure}

\begin{figure}
	\resizebox{\hsize}{!}{\includegraphics{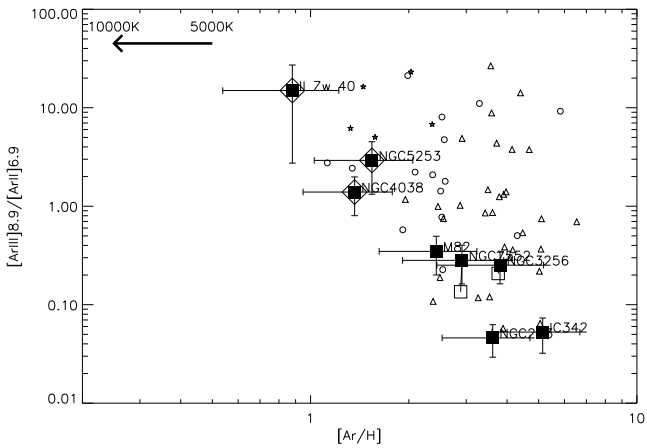}}
	\caption{Excitation against abundance for
	argon. Symbols are as in Fig.~\ref{neab}.
	\label{arab}}
\end{figure}

\begin{figure*}
\centering
	\includegraphics[width=8.5cm]{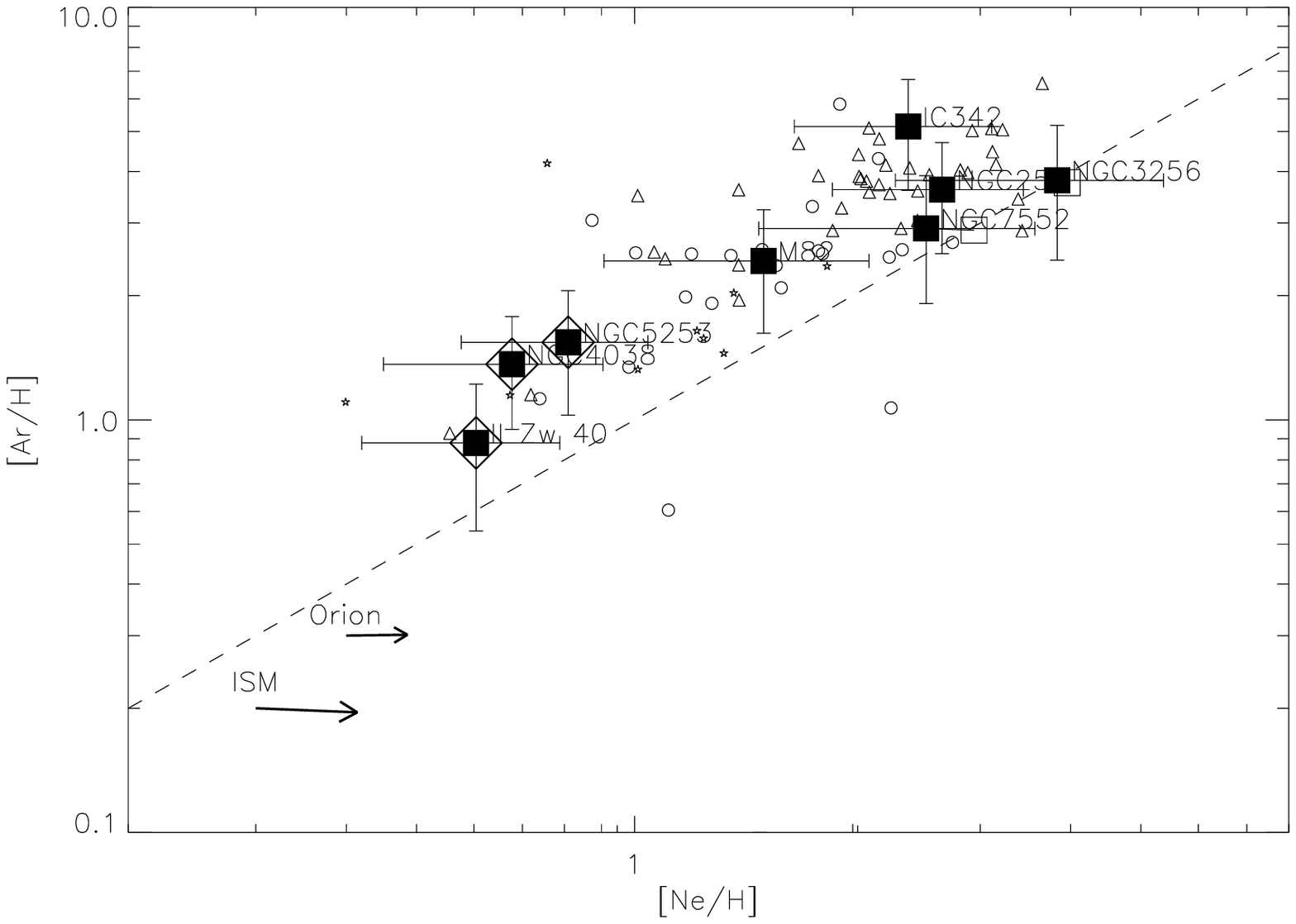}
	\includegraphics[width=8.5cm]{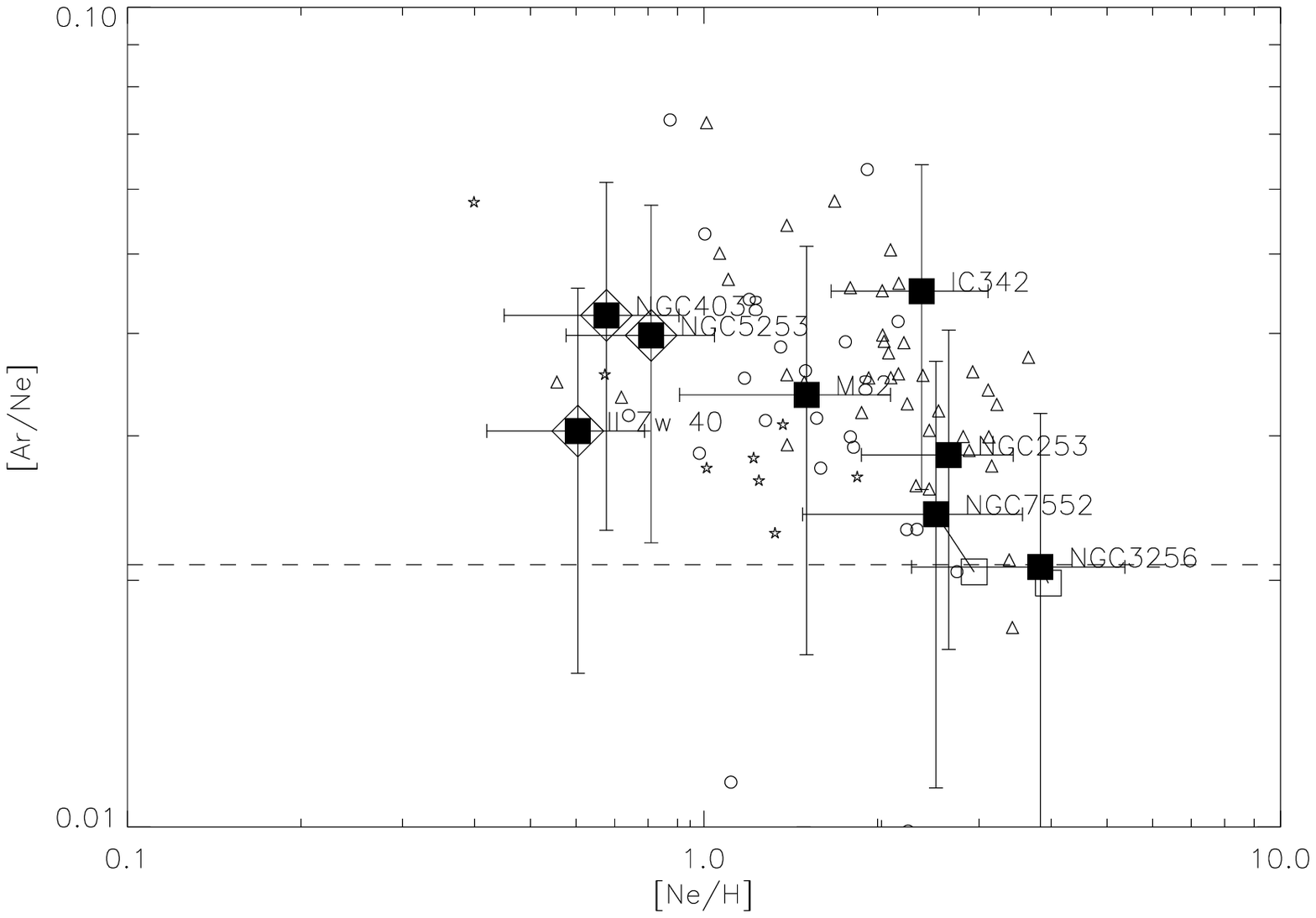}
	\caption{
	{\em Left: }[Ne/H] abundance vs. [Ar/H] abundance. The key is as 
        in Fig. \ref{nearex}. 
	{\em Right: }[Ar/Ne] abundance ratio vs. [Ne/H] abundance. The key 
        is as in Fig. \ref{nearex}.
	In both panels, a dashed line represents the solar abundances and 
        abundance ratios for these elements.  The arrows marked `ISM'
	and `Orion' denote the effect on the data of using different reference
	abundances other than solar - `ISM' interstellar medium from
	\citet{wilms00} and Orion abundances from \citet{sim98}. 
	Adopting these different
	reference abundances only change the absolute values but not
	the correlation.
	\label{nearab}}
\end{figure*}

In Figs. \ref{neab} and \ref{arab} we show the relationship between
excitation and abundance\footnote{We calculate 
our abundances relative to solar \citep{gre98} to
be consistent with extragalactic abundance work. 
It is widely thought that the Sun is under-abundant in heavy metals
compared to its local neighbourhood \citep[e.g.,][]{sno96}. Adopting different
reference abundances will only change the absolute values but not affect
the correlations discussed in the following. For information,
we show in Figs. \ref{nearab}-\ref{arsab} the effects of different
reference abundances by arrows marked `ISM'  
\citep[interstellar medium][]{wilms00} and `Orion' \citep{sim98}} for Ne and Ar 
for the starburst sample\footnote{For the starburst regions A and B/C 
in NGC 3690, argon abundances are not presented since no measurement
exists for the strong $[ArII]6.99\mu m$ line. Therefore any
calculated abundances for these regions would be severe
underestimations.}.  
A trend of decreasing excitation with increasing
abundance is clearly seen in Ne and Ar for the starburst regions. The \wr\ galaxies 
have the highest excitation and
the lowest abundance. In general, the non-\wr\ starbursts in our sample have
super-solar metallicities $(X/H)\sim 1-3 (X/H)_{\sun}$ (where X is Ne or Ar) 
whereas the \wr\ and BCD galaxies are sub-solar.  
As discussed in Sect. \ref{et}, our adopted `representative' electron 
temperature of 5000\,K is
a good assumption for high metallicity starbursts and high metallicity 
\hii\ regions.  However, the same is not true for the BCD galaxies and 
the lower metallicity \hii\ regions among the sample of \citet{giv02}. 
Therefore, the effect of changing 
the electron temperature from 5000 to 10000\,K
on the calculated abundances is shown with 
an arrow in the diagrams. 
By adopting a more appropriate electron temperature, 
the absolute infrared-derived abundances of the low metallicity BCDs are 
close to the optically-derived values [$0.2(X/H)_{\sun}$] 
\citep[e.g.][]{hun82,vac92,wal93,mas94}. Overall, the effect 
on the excitation/abundance diagram of Fig. \ref{neab} 
of adopting individual electron temperatures (which are unavailable
for the entire sample) for each starburst would be to
tilt the correlation: the high excitation / low metallicity end would
move to $\sim$2 times lower abundance, while the high metallicity/low
excitation end remains unchanged. 

From the literature we found accurate electron temperatures for the \ir\
starburst regions of M82, \n253, II Zw 40
and \n5253 (as described in Sect. \ref{et}) which we used to
re-calculate 
the elemental abundance of neon (Table \ref{tabab2}) in these galaxies.  
II Zw 40 and \n5253 ($T_{e} \sim 12500K$ and $\sim 11000K$, respectively) lie at the low
abundance, high excitation end of our abundance-excitation correlation,
while M82 and NGC\,253 ($T_{e} \sim 5000K$) lie at the opposite extremity.  We then 
defined a linear relationship between excitation and true 
Ne abundance (calculated using the measured actual
electron temperatures) defined by these two pairs
of galaxies.  We could then estimate neon abundances for
the remaining galaxies in the sample by interpolating along this
relation based upon their \neiii 15.5/\neii 12.8  excitation ratios.  This process yielded 
abundance estimates that are essentially independent of our knowledge
(or lack thereof) the
true electron temperature for each source.  The
estimated abundances (Table \ref{tabab2}) 
for the low excitation sources are all close to $\sim 2Z_{\sun}$.
 
\begin{table}
\caption{Neon abundances for our sample of galaxies calculated assuming a fixed
electron temperature of 5000K (column 2) and individual electron
temperatures (3rd column).  For the latter, abundances are
derived from interpolating between objects with known electron
temperatures (values indicated by bold face), on the basis of
their \neiii 15.5/\neii 12.8 excitation ratios. 
\label{tabab2}}
\begin{tabular}{lll}
\hline\hline
Name			&	$[Ne/H](5000K)$	&	$[Ne/H](T_{e})$\\
\hline
             IC342	&	2.4			&	2.1\\
          II Zw 40	&	0.6			&	{\bf 0.3}\\
               M82	&	1.5			&	{\bf 1.5}\\
            NGC253	&	2.7			&	{\bf 2.7}\\
           NGC3256	&	3.8			&	2.1\\
          NGC3690A	&	2.2			&	2.0\\
        NGC3690B/C	&	1.5			&	1.9\\
           NGC4038	&	0.7			&	1.9\\
           NGC4945	&	1.5			&	2.1\\
           NGC5236	&	3.1			&	2.1\\
           NGC5253	&	0.8			&	{\bf 0.4}\\
           NGC7552	&	2.5			&	2.1\\
\hline
\end{tabular}
\end{table}

As `primary' products of stellar nucleosynthesis, the 
abundances of neon and argon are expected to be correlated. 
The overall abundance
relation between Ne and Ar is shown in Fig. \ref{nearab} which indeed shows
a clear correlation between the two elements that holds for all objects
in our sample, from the BCDs and WR galaxies to the dusty starbursts. 
This implies that the ISM of these
galaxies has been enriched by similar processes for both elements.
The [Ar/Ne] ratio shown in the right panel (for which the $T_e$ effects
for the two elements largely cancel) suggests an above solar average 
argon to neon abundance ratio with no obvious trend among the galaxies.

\subsubsection{The abundance of sulphur}
\label{sulplot}

\begin{figure}
	\resizebox{\hsize}{!}{\includegraphics{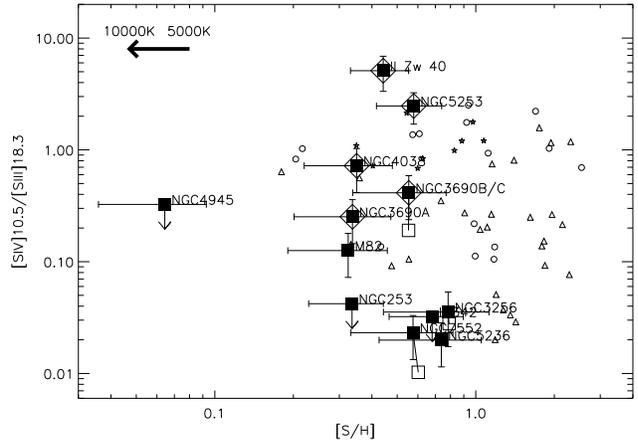}}
	\caption{Excitation against abundance for
	Sulphur. Symbols are as in Fig. \ref{neab}. Unlike the corresponding
        figures for neon and argon, no correlation is seen.
	\label{sab}}
\end{figure}

We discuss the abundance of sulphur separately since we do not see the clear correlation between excitation
and abundance that is seen in Ne and Ar for the starbursts
(Fig. \ref{sab}). In addition, {\em all} galaxies appear to have
sub-solar sulphur abundances, with \n4945 being anomalously low. 
Furthermore, on comparison to the abundances of Ne and Ar, we find that the galaxies
not showing \wr\ features have lower 
S abundances ($<[S/H]>=0.6$) than
expected from their typically high Ne and Ar abundances 
($<[Ne/H]>=2.4,\ <[Ar/H]>=2.7$).  
Yet the S abundances for the \wr\ galaxies (which typically have 
lower Ne and Ar abundances) are similar to that expected
from their Ne and Ar abundances ($[X/H]\sim0.2$).

\begin{figure*}
\centering
	\includegraphics[width=8.5cm]{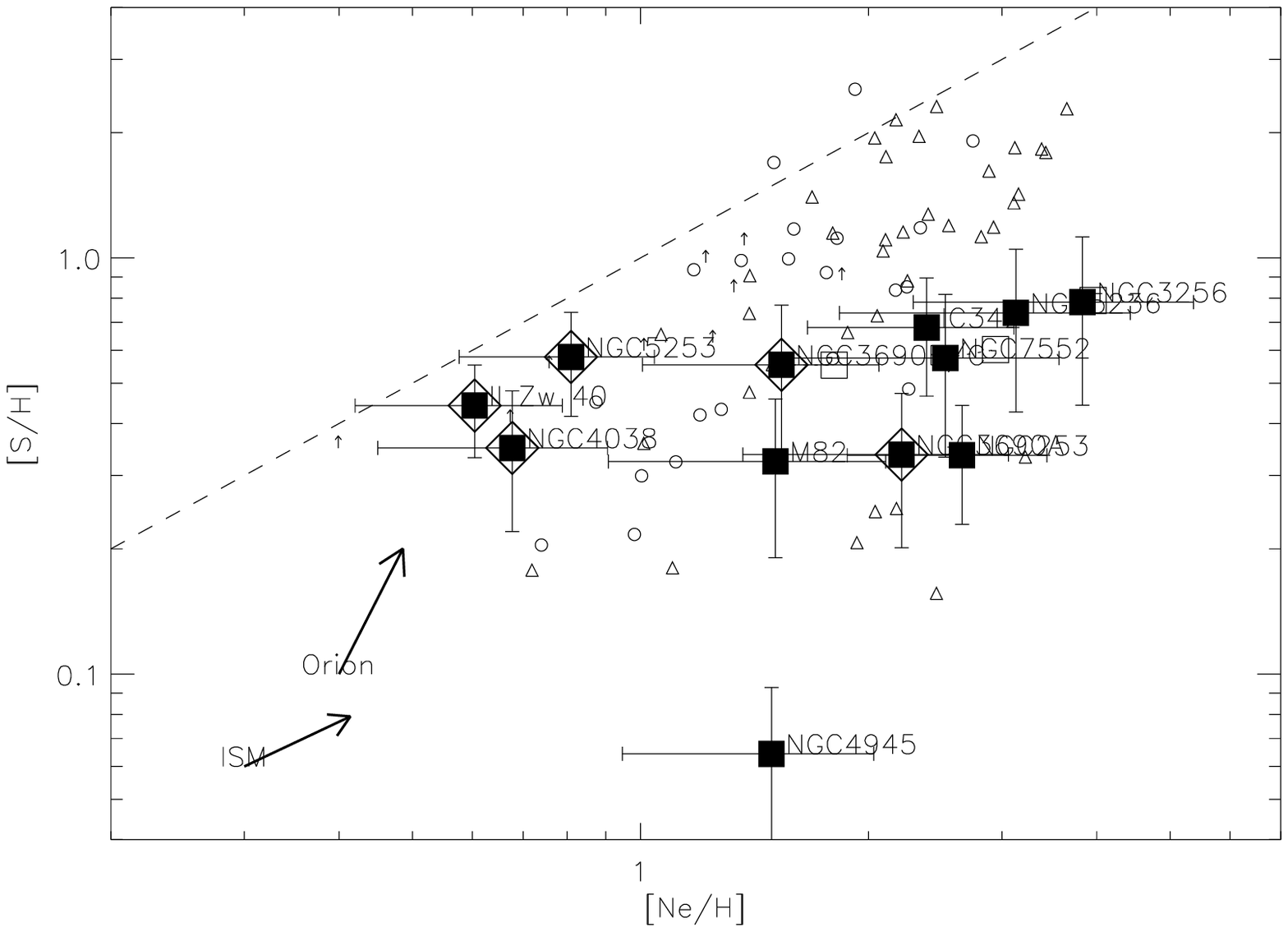}
	\includegraphics[width=8.5cm]{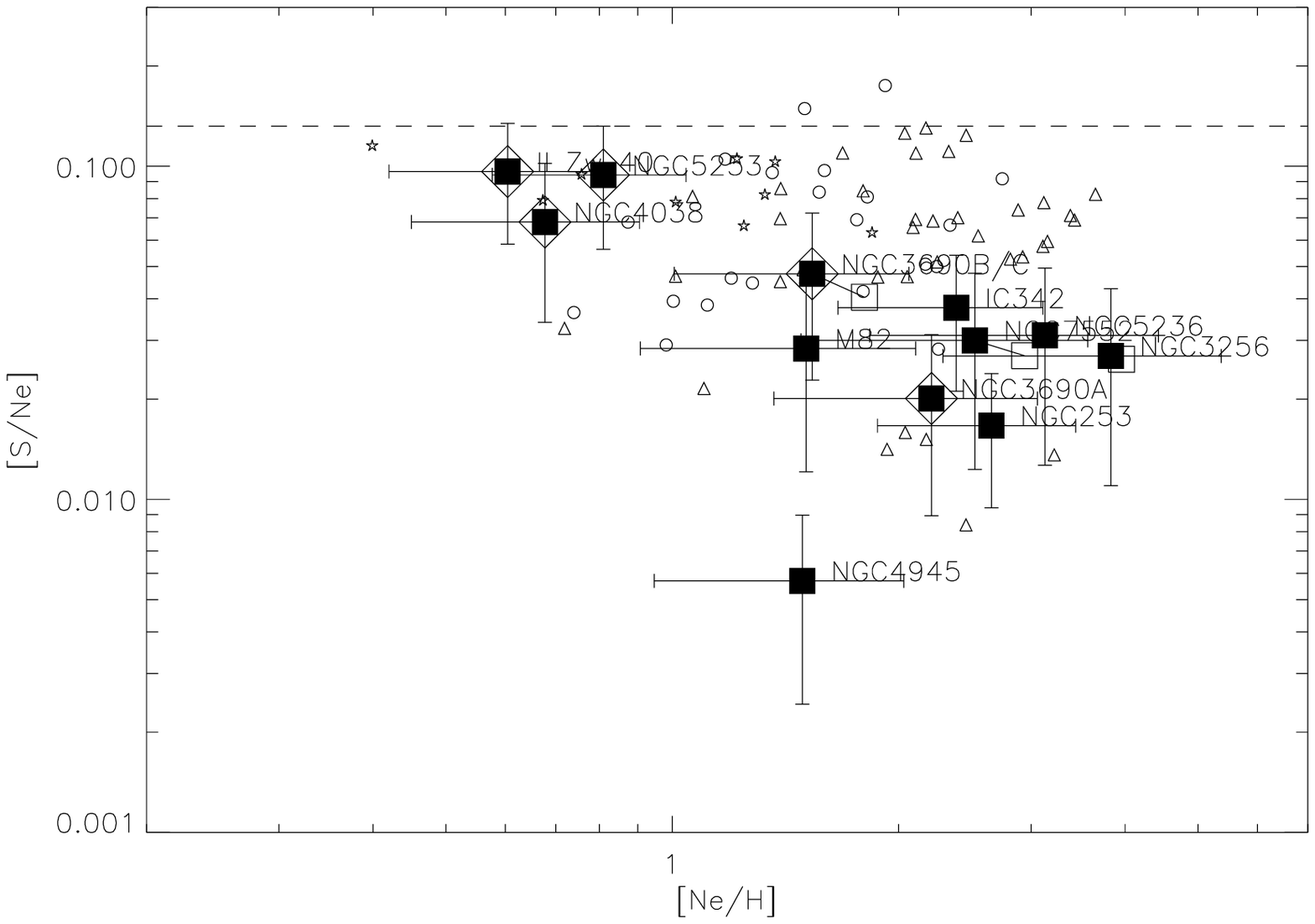}
	\caption{{\em Left: }[Ne/H] vs. [S/H] abundance. The key is as in 
        Fig.~\ref{nearex}.  
	{\em Right: }[S/Ne] v [Ne/H] abundance. The key is as in Fig.
	\ref{nearex}. 
        For both panels, dashed lines represents the solar abundance and
        abundance ratios. 
	\label{nesab}
}
\end{figure*}

\begin{figure*}
	\includegraphics[width=8.5cm]{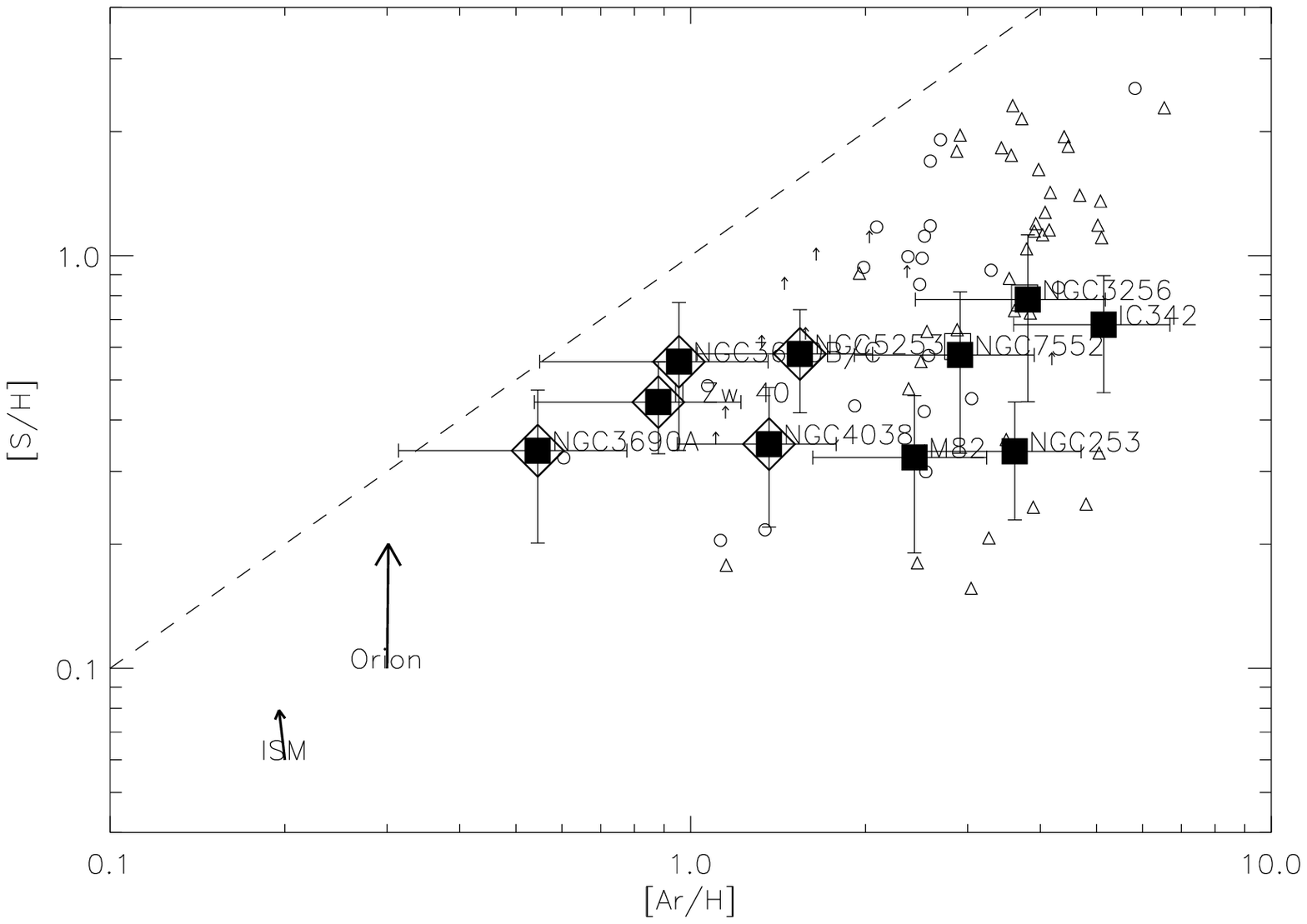}
	\includegraphics[width=8.5cm]{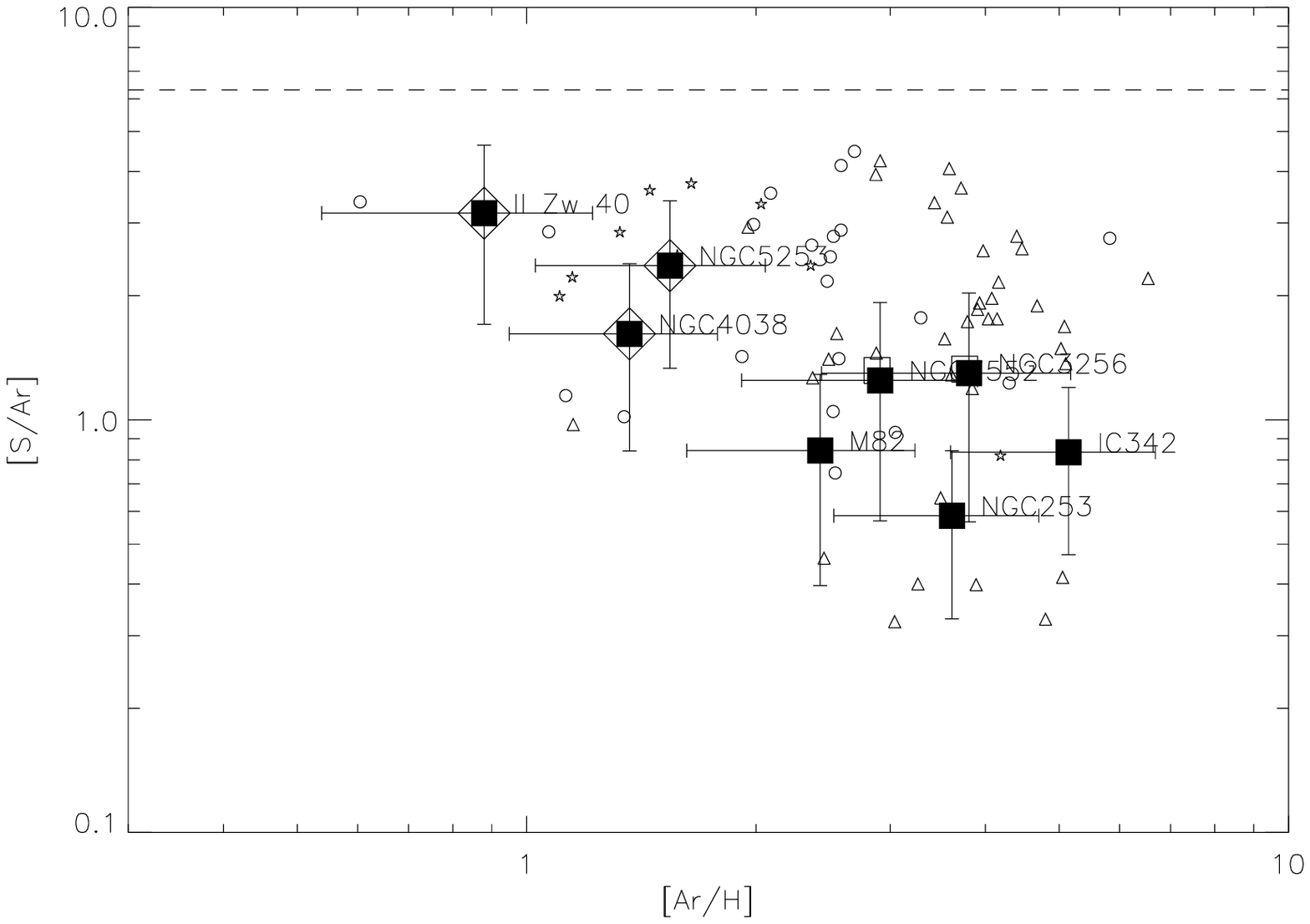}
	\caption{{\em Left: }[Ar/H] v [S/H] abundance. The key is as in Fig.
	\ref{nearex}. 
	{\em Right: }[S/Ar] v [Ar/H] abundance. The key is as in Fig.
	\ref{nearex}. 
        For both panels, dashed lines represents the solar abundance and
	abundance ratios.
	\label{arsab}
}
\end{figure*}

We confirm this result in Figs. \ref{nesab} and \ref{arsab}, where we
investigate the correlations between S abundances and Ne and Ar.  
S is also 
expected to be correlated with Ne and Ar since they are all
primary products of stellar nucleosynthesis.  
However, we see no
correlation between the abundances of S in the galaxies with either 
Ne or Ar. 
The right panels of 
Figs. \ref{nesab} and \ref{arsab} compare the abundance ratios [S/Ne] and
[S/Ar] with the Ne and Ar abundance. They reflect the trend towards
a stronger sulphur `deficiency' at increased Ne or Ar abundance noted above.
S is often referred to as an excellent tracer of metallicity
\citep[e.g.][]{oey00} and in studies of the ISM is commonly used as a
non- or slightly-depleted ($\la 30\%$) element 
\citep[e.g.][]{wilms00}. However, the significant underabundance of 
S implied by the results of this study clearly
has implications for the use and interpretation of sulphur \fsl s
and sulphur abundances in starburst sources. This result suggests that
S is {\it not} a good tracer of metallicity in starburst galaxies.
We discuss possible explanations for this depletion of sulphur below.

\section{Discussion}

\subsection{Comparison to optical/NIR derived abundances}

A direct comparison to the abundances derived from optical/NIR studies is far 
from straightforward.  
In particular, our obscured starbursts are not directly comparable to 
optical derived abundances since the 
regions we are probing may be completely obscured.  
Moreover, when making comparisons 
for our sources of large angular size one must ensure that the spectra in 
both regimes are probing the same regions.  

The abundances obtained from our \ir\ study  
complement optical/NIR determinations by adding Ne, Ar and S which 
have predominant excitation species in the \mir\ wavelength range.  
We searched the 
literature for abundances determined from optical spectroscopy 
for these elements, or for oxygen with which they are
expected to be correlated.
For II Zw 40 and NGC 5253, optical based abundances 
are consistent with our \mir\ results \citep[$Z \sim 0.2Z_{\sun}$][]{kob97}.  
For our low excitation
starbursts with metallicities close to $2Z_{\sun}$, while optically
derived values are in agreement that the starbursts have super-solar
metallicities, the estimates are slightly discrepant with our \ir\
determinations. For NGC 3256 \citep[][ and references therein]{bre02}
and NGC 7552 \citep{sto94} the derived metallicities are slightly lower
$Z\sim1.5Z_{\sun}$, whereas for N5236 the determinations is slightly
higher $Z\sim2.6Z_{\sun}$ \citep{sto94}.

\subsection{Excitation and Abundance} 

The strong negative correlations seen for Ne and Ar in Sect. \ref{near}
indicate that `spectral hardness' of the ionising stars 
is inversely
correlated with metallicity. This correlation
is likely to arise from several factors, such as: (a) The influence of metallicity on stellar atmospheres and the 
     resulting stellar spectral energy distribution, for a given effective 
     temperature of the star. (b) Variations of stellar evolutionary tracks
     with metallicity. (c) Variations of IMF with metallicity. The observed
     {\it spread} of the correlation could result from variations in the
     electron temperature, starburst age and the IMF among sources in our 
     sample. We discuss the causes of both the correlation and the scatter
     in this section.

Theoretical models of hot star atmospheres indicate 
that metallicity variations 
affect stellar wind
strengths and, therefore mass-loss rates, as a result of changes 
in the opacity in the line-driven winds.  Changes in metallicity also 
induce changes in the 
the effects of
line blocking and blanketing.
\citet{tho00} and \citet{giv02} present a 
    discussion of the effects on stellar populations. \citeauthor{giv02}
    find that the increase in excitation of Galactic \hii\ regions with
    galactocentric radius cannot be due to this effect alone [(a) in the list 
    above]. Increasing effective temperatures must strongly contribute, in
    addition to the hardening of stellar SEDs for fixed $T_{eff}$. 
	The same can
    be concluded for our galaxies by comparing the observed spread in 
    excitations to the \citeauthor{giv02} models.

The effective temperature of a star-forming cluster is determined by
{\em at least} three related factors: age, IMF, and metallicity. 
While we cannot differentiate
   between the individual effects of these three parameters on our data for 
   single objects, we can discuss the influence they may have on our 
   abundance-excitation correlations. 
Stellar evolutionary tracks across the H-R diagram also clearly display a 
dependence on metallicity
\citep[e.g.,][]{scha92}. For example, lower metallicity tracks have a
`hotter' main
sequence. This explains why high excitation regions are 
found primarily at low metallicity \citep[as Z varies from
$Z_{\sun}$ to $0.2Z_{\sun}$ the observed Ne excitation ratio increases by
factors $\sim4-10$,][]{tho00}.
Evidence for the Galactic Centre may suggest this effect (b) is even 
stronger \citep[see discussion in][]{tho00}.

The question of whether (c) the IMF may vary among clusters of different metallicities
remains open.  Following arguments as in \citet{tho00}
we consider such IMF variations a minor contributor to our correlations.
They showed that effects of moderate IMF slope variations on the excitation
are not large, and discussed evidence in favour of the most massive stars
being present in the initial mass function of objects for a wide range of 
metallicities.

We have already noted that some of our sources (particularly the BCDs) 
have measured electron temperatures
that differ from the representative value of $T_{e}^{R}=5000K$.  
However, it seems unlikely that variations in $T_{e}$ could cause a 
correlation between `hardness' (effectively $T_{eff}$) and metallicity, 
as the dependence of abundance on electron temperature is weak 
\citep[$T_e^{-0.7}$][]{giv02}.
If $T_e$ increases with
metallicity, as is suggested by the electron 
temperatures we found for NGC 253, M82, II Zw 40 and NGC 5253, then
the variations in $T_{e}$ for our sample of galaxies
will affect the slope of the correlation we find between `hardness' and metallicity, 
but will not destroy it. 
In the unlikely case that $T_{e}$ does not systematically vary with
metallicity, then 
the variations in $T_e$ in our 
sample will serve only to increase the
spread in the correlation.

The observed scatter in the correlation may also result from
differences
in both starburst ages \citep[as argued by e.g.][]{tho00,giv02} 
and the IMF for different star-forming clusters.

\subsection{Comparison to \hii\ regions}
We also note from our
`hardness'-metallicity plots, that for a given metallicity, the starbursts
have relatively lower excitation than the \hii\ regions.  Similarly
low excitations have been noted in previous studies of extragalactic star-forming regions
\citep{doy94,doh95,tho00}.
Our improved empirical
database of starbursts (this paper) and \hii\ regions \citep{giv02}, 
reinforces this result.

The low excitation of galaxies has been attributed to low
upper-mass-cutoffs ($< 30M_{\sun}$) and ageing effects \citep{tho00}.
The starbursts in our sample represent complex starburst systems with 
{\it possibly} multiple stellar populations lying within the SWS aperture, rather than 
the single
stellar populations that comprise \hii\ regions.  This difference may
account for the low excitation since the SWS aperture may include emission not only from
the `youngest burst' population but also 
from older stars within the host galaxy.  The net effect
of this `dilution' is to lower the {\em mean effective temperature} of 
the starburst, which will result in the starbursts having lower `hardness' ratios
\citep[cf Fig. 5 in][]{mar02b}. This lowering of the {\em mean effective
temperature} may mirror the effect of an ageing
stellar population \citep[as predicted by][]{tho00} on the observed line ratios.

\subsection{Weak sulphur lines and the origin of low sulphur abundance}
\label{sulph}

We have shown in Sect. \ref{sulplot} that the gas phase sulphur abundances 
do not follow the consistent behaviour seen in neon and argon. With the
exception of the low metallicity systems, the measured S abundance is 
consistently smaller than that of Ne or Ar, by up to an order of 
magnitude in the metal-rich and dusty objects. In addition to the 
question of the origin of this deficit, the weakness of the S lines 
has practical 
implications for future observations of dusty starbursts both locally and at 
high redshift: sulphur lines are less favoured in comparison to the
[Ne\,II] line
as tracers of starburst emission than one
would assume from nebular models. Instead of photoionisation models,
empirical templates may be the tool of choice for flux
predictions. \citet{gen98} showed that low redshift ultraluminous
infrared  galaxies follow the trend of weak 
S lines relative to Ne, with Arp 220 being the most pronounced example.

Previous studies of Galactic H\,II regions also found low S abundances.
From infrared spectroscopy, \citet{sim90}, \citet{sim95} and \citet{sim98}
derived S and Ne abundances corresponding, on average, to a subsolar ratio of 
S/Ne, and suggested depletion of sulphur onto grains as a possible cause. 
This S deficiency was recently put on a much firmer observational basis by
\citet{mar02}\footnote{see also Fig. \ref{sab} where the
under-abundance of S found by \citet{mar02} is seen if one applies the
offset marked by an arrow due to the differences in adopted electron temperature},
who used ISO observations of a larger sample to discuss Galactic abundance
trends. Compared to the Ne and Ar abundances (with mean values
similar to solar), their derived
sulphur abundances are more scattered 
and are lower than solar by a factor $\la$3. 
For individual, higher-than-solar-metallicity extragalactic H\,II regions,
\citet{dia91} suggested a low S/O abundance ratio based on optical ($<1\mu$m)
spectroscopy, but they emphasised the difficulty of determining abundances
for high metallicity regions using optical data. These difficulties
also apply
to a possible Ar/O deficit at high metallicity \citep{gar02}, 
and are further reflected in the fact that other
optical studies do not suggest a low S/O or Ar/O ratio for super-solar oxygen
abundances \citep{vanzee98}.

For the low S/H abundances measured in Galactic \hii\ regions,
\citet{mar02} propose an explanation involving uncertainties in electron temperature
and density. While such uncertainties clearly exist, two arguments suggest
that they are of secondary importance for the galaxies in our sample.
First, the density measured from the [S\,III] lines is low for all galaxies
in our sample and does not show the occasional high values that call the applicability 
of the low density approximation for determination of the S abundance in
compact \hii\ regions into question. Second, in addition to the inherently small temperature
sensitivity of abundance determinations from fine structure lines, the 
remaining abundance variations caused by variations in temperature are 
expected to correlate among the three elements S, Ar, and Ne. 
Varying the adopted electron temperature from 5000 to 12500\,K 
decreases the inferred abundances by factors of 2 to 2.5 for all three
elements simultaneously.
Large temperature uncertainties therefore should not selectively lower the 
sulphur abundance and increase the scatter in excitation versus
abundance for sulphur alone, but rather affect all 
three
elements similarly. In contrast, observations single out sulphur as being 
peculiar both in our data and in the H\,II region data of \citet{mar02}
(their Figs. 18 and 19).

Another issue relevant to the question of S abundance is the
ionisation correction factor used to compute
the total sulphur abundance.
Since the S underabundance occurs in {\em low} excitation objects, 
the missing fraction are unlikely to exist in
higher ionisation stages (i.e. beyond the 47eV ionisation potential of 
$S^{3+}$) but more realistically in $S^+$ (23eV). 
However, the ICF calculations of \citet{mar02}
show this effect to be small for the excitation range of our
objects. 

We consider a direct nucleosynthesis explanation of the sulphur deficiency
to be unlikely. Neon, argon and sulphur are all primary elements expected to trace 
each other reasonably well. Variations will be related to
the different stellar mass ranges contributing to the yields for each
element. Extreme assumptions for the initial mass function, and/or 
variations of stellar lifetimes with mass, can
influence the ratio of neon and sulphur abundances, since these
two elements are produced mostly in supernovae with high (neon) and lower mass
(sulphur) progenitors, respectively  \citep{woo95}. However, argon is produced over 
a very similar mass range as sulphur. Therefore, the observed
abundance 
pattern of both neon and argon being `normal', but sulphur 
being anomalously low by a large factor cannot easily be explained by
differences in the mass ranges of the SNe.

The calculated elemental abundances are also heavily dependent upon
the atomic data and therefore errors in the published values might be
responsible for erroneous sulphur abundances. However, we have
used the most recently determined collision strengths for 
both doubly and triply
ionised S from \citet{tay99} and \citet{sar99}, respectively.  These
papers use improved and more complex methods to determine accurate
collision strengths than previous, commonly used studies 
\citep[e.g.][]{galv95,joh86}. While we identify atomic data as a
possible contributor to the sulphur underabundance problem, the
abundances we present are correct with respect to the most up-to-date
atomic data available. 

A consistent explanation of both extragalactic and Galactic observations
would be a tendency of sulphur to be more strongly depleted onto dust than 
the noble gases neon and argon. 
However, ultraviolet absorption measurements in the diffuse 
interstellar medium often suggest most sulphur exists in the gas phase
\citep{sav96} with low depletion 
\citep[$\sim 30\%$][]{whi84,wilms00}. 
Other ultraviolet ISM studies suggest a correlation between sulphur 
depletion and mean
hydrogen density \citep{gon85,har86,vanste88}, with depletions reaching
a factor of 10 for lines of sight with mean hydrogen densities $\sim 1cm^{-3}$. 
For molecular regions, strong sulphur depletion is predicted 
by chemical models \citep[e.g.][]{dul80,pra82,ruf99}. 
Unfortunately little is known about the H\,II
regions probed by our observations beyond the infrared studies cited above. 
While depletion is a natural
explanation of the failure of the S abundance data to trace the noble gas
abundances, a quantitative understanding under the conditions of our
higher than solar metallicity H\,II regions remains to be obtained.
Clearly, depletion would have to depend on ISM conditions to explain both
the low metallicity dwarfs (with no S deficit both from our results and
numerous optical studies) and the higher metallicity starbursts. 
While noting these open questions, we consider depletion of sulphur
onto grains the best explanation for our observations.

\subsection{Wolf-Rayet Galaxies}
\label{wrsect}

\wr\ galaxies exhibit signatures of \wr\ stars in their integrated visible
spectra and have been detected in a variety of galaxy types (e.g. BCDs,
H\,II galaxies, AGN, LINERs etc.) but not in all objects of a particular 
class \citep{scha99}.
In general, the detection of \wr\ signatures in a galaxy has been
interpreted as an indicator of young burst age ($3 \la t_{sb} \la 8Myr$) and the
presence of massive stars ($M_{upp}\ga20M_{\sun}$), since the progenitors 
of \wr\ stars are postulated to be
massive O stars.  The \wr\ stage is thought to appear for a short time 
during the evolution of a simple stellar population. 

In our sample, \wr\ galaxies are clearly separated from non-\wr\ 
galaxies.
\wr\ galaxies generally have higher
excitation and lower abundances than non-\wr\ objects.
The current inventory of \wr\ features in external galaxies \citep{scha99}
is definitely not unbiased, but in general \wr\ signatures have 
been detected mostly in low metallicity systems. 
The low detection rate of \wr\ stars in high metallicity systems is contrary
to expectations from stellar evolution theory
\citep[e.g.][]{mey95}.
\citet{lei95} for example,
model the ratio of \wr\ stars to O stars in evolving starbursts using the Geneva tracks, and derive an order of magnitude increase in the 
metallicity range 0.25 to 2 $Z_{\sun}$; this range approximately corresponds to the 
metallicity spread in our sample.

Recent surveys of a large number of \hii\ regions \citep{bre02,cas02,pin02}
located in the {\em disks} of local
high metallicity galaxies have revealed the presence of weak \wr-like features
in $\sim30\%$ of the surveyed regions, which implies that
\wr\ stars do in fact exist in high metallicity environments.  This result
confirms previous work on smaller samples \citep{gus00,scha00} and
high metallicity systems included in other \wr\ samples
\citep[e.g. NGC\,3049;][]{vac92}. 

Yet, in highly obscured star forming regions, the presence of \wr\
stars remains to be confirmed. 
Our higher metallicity,
non-\wr\ starbursts 
are all well studied and strong \wr\ features would not have escaped detection.
\n5236 is the only exception. Despite its location in the 
excitation and abundance 
plots that is clearly consistent with the 
non-\wr\ starbursts, \citet{bre02} have recently
identified possible \wr\ features in the optical spectrum of a nuclear
hot-spot (hot-spot M83-A in their paper) which lies within our
\sws\ aperture.  The red and blue spectra they present have relatively low signal-to-noise and show no 
C\,IV$\lambda$5808 emission but do show a weak `blue-bump' around
4650\AA\ which is indicative of the presence of WN
stars.  Nevertheless, it is unlikely that the blue
bump is solely due to WN stars. \n5236 has a N\,III $\lambda$4640 / He\,II $\lambda$4686 line
ratio that is greater than unity which \citet{schm99} have shown 
cannot be reproduced by mixtures of known \wr\
stellar spectra. \citeauthor{schm99} suggest that a large contribution
from Of stars could also produce such emission features.
Similar line ratios have also
been found in other high metallicity systems \citep[see][]{scha00}.
  
The separation of NGC 5236 with respect to II Zw 40 and NGC 5253
is not unexpected since they represent two very different classes of
object.  NGC 5236 is large barred spiral galaxy which is actively
forming stars on kiloparsec scales, whereas for II Zw 40 and NGC 5253 the
majority of star-formation and IR emission is thought to originate
in 1 to 4 super-star clusters \citep[c.f.][]{gor01,beck02}. Thus its
location amongst the non-BCD galaxies is reasonable.

There could be several reasons for the non-detection of \wr\ features in
most of our 
high metallicity objects: a special star formation
history (no significant part of the population currently being in the
brief \wr\ phase); dust obscuration of the regions actually
hosting WR stars; and/or changes in spectral
signatures of \wr\ stars at higher metallicity.   

We do not consider fine-tuned star formation histories to be the dominant
reason for the dichotomy in our sample. It seems very unlikely that only
metal-rich objects are observed at extremely young or late ages (and
with O stars still present to power the starburst).
Circumventing this problem by postulating that the metal-rich objects are
fully self-enriched in the current (almost terminated) burst would similarly
require stringent assumptions on timing and the enrichment process. 
Most importantly, attempts using all available constraints for a detailed 
reconstruction of the star formation history in M\,82, the prototype 
non-\wr\ object in our sample, predict a major part of the burst
population to be in the `\wr\ phase' \citep{rie93,nfs02}. 

Obscuration of the regions hosting \wr\ stars is likely to be a
significant effect.  The presence of \wr\ stars is indicated by 
weak stellar emission features around 4650\AA.
In several
of our objects, the most active star forming regions are
considerably dust-obscured, and in the absence of convincing near or
mid-infrared \wr\ tracers \citep{lum94}, 
detection of any `classical' optical \wr\ features 
is difficult. 
This is especially true if other, less obscured
regions dilute the blue part of the visible spectrum. This is the case 
for M\,82 which has a strong post-starburst component in its optical 
spectrum \citep[e.g.][]{ken92}.  
It is worth noting that the high metallicity \hii\ regions within
which optical \wr\ features have been detected \citep{bre02,cas02,pin02}
are located within the disks of their host galaxies, 
and may suffer less extinction (and dilution) 
than our deeply obscured \ir\ star forming regions.

The role of \wr\ stars in metal rich starbursts also depends on still
uncertain elements of the post main sequence evolution of massive
stars. This is true for both the details of the evolutionary tracks,
and the assignment of spectra to those tracks. For example, the 
ionising spectra of \wr\ stars may not be as hard as commonly assumed; 
this has implications for the spectra of composite
starburst populations \citep{bre99,cro99,bre02}.
Additionally, evidence suggests that \wr\ stars in higher 
metallicity environments display different spectral signatures and
weaker lines in the 
visible, compared to those in lower metallicity environments 
\citep{masj98}. Recent modelling by \citet{bre02} shows that 
\wr\ photons have a
negligible contribution to the nebular ionisation as they are absorbed
in the stellar atmospheres. The presence of \wr\ stars in our galaxies
could be reflective only of young age.
Moreover, in the Galactic centre \citet{tho00} found that the 
majority of ionising flux originates from fairly cool (20000 to 30000\,K)
supergiants rather than the main sequence and hot \wr\ stars expected from
a direct interpretation of the Geneva tracks. All this warns that a 
direct transfer from theoretically predicted \wr\ star numbers to observed signatures in
high metallicity environments is currently difficult.

We note that the dichotomy 
between \wr\ and non-\wr\ objects in our
sample could also be enhanced by an overestimation of the role of \wr\ stars
in the {\em low metallicity} objects. Recent studies suggest the existence in
very young star forming regions of non-\wr\ (core hydrogen burning) stars 
that, nevertheless,
show \wr-like broad emission \citep{mas98}. This might 
additionally enhance the incidence of `\wr\ features' among
the very youngest, highest excitation objects \citep{schm99}.

\section{Conclusions}

We have presented \mir\ spectral line data from a spectroscopic survey of a sample of
starburst galaxies as seen by \sws.  These data 
can be used as a reference database for comparison to
future observations of star-forming galaxies with infrared telescopes such
as {\em SIRTF}, {\em SOFIA} and {\em Herschel}.

From this database we have investigated the excitation and
abundances of the sample using fine structure lines
of the primary nucleosynthetic products neon, argon and sulphur and 
the H-recombination lines Brackett
$\alpha$ and Brackett $\beta$.
Abundance and excitation are inversely correlated for Ne
and Ar.  In addition, a comparison to local \hii\ regions shows that
{\em for
a given metallicity starbursts are of relatively lower excitation than
the \hii\ regions}. The excitation of mid-infrared starburst spectra is 
hence governed by a combination of metallicity and other effects like ageing
of the population, which should be accounted for in modelling starbursts as
composite \hii\ regions.  

An analysis of the excitation and abundance as traced by fine structure
lines of of sulphur indicates that sulphur is approximately 3 times
underabundant for the low excitation metal-rich galaxies relative
to Ne and Ar with which it should be
correlated since it is also a primary product of nucleosynthesis. 
We favour depletion onto dust grains as the most likely cause of this 
relative underabundance.
The weakness of the sulphur lines may favour neon as an indicator of
star formation in future infrared spectroscopy of faint galaxies. 
In addition the derived low sulphur abundances imply that S is 
not a good tracer of metallicity.

Our sample displays a dichotomy between galaxies showing \wr\ features
in the optical, which are of high excitation and low metallicity, and those
without \wr\ features, which are of low excitation and high metallicity. 
This is opposed to the expectation of higher \wr\ fractions at
higher metallicity from stellar evolutionary models. The most plausible reasons
for this behaviour include obscuration coupled with a lack of convincing
\mir\ \wr\ tracers, dilution by less obscured regions in the optical
spectrum and changes in the spectral signatures of \wr\ stars at
higher metallicity.

The excitation probed by fine structure lines reported in this paper
have been combined with those from a sample of active galaxies
and are presented in a
related paper \citep{stu02}. \citeauthor{stu02} have constructed infrared analogues to the
classical optical excitation diagnostics of \citet{vei87}.  The results show a
clear separation between the two populations and the new excitation
diagnostic can be used to identify powering mechanisms of obscured
sources detected in future infrared surveys.


\begin{acknowledgements}
We would like to thank the referee Dr M Sauvage for his constructive
and useful comments. 
The ISO Spectrometer Data Center at MPE is supported by DLR under grant
50 QI 0202.  This research is partly supported by the
German-Israeli Foundation (grant I-0551-186.07/97). 
\end{acknowledgements}


\bibliographystyle{aa}

\end{document}